\PassOptionsToPackage{prologue,dvipsnames,svgnames}{xcolor}
\documentclass[lettersize,journal]{IEEEtran}
\usepackage{amsmath,amsfonts,amssymb}
\usepackage{algorithmic}
\usepackage{algorithm}
\usepackage{array}
\usepackage[caption=false,font=normalsize,labelfont=sf,textfont=sf]{subfig}
\usepackage{textcomp}
\usepackage{stfloats}
\usepackage{url}
\usepackage{verbatim}
\usepackage{graphicx}
\usepackage{cite}
\usepackage{amsthm}
\usepackage{float}
\usepackage[dvipsnames,svgnames]{xcolor}
\usepackage{kotex}
\usepackage{tikz}
\usepackage{pgfplots}
\usepackage{adjustbox}
\usepackage{multirow}
\usepackage{grffile}
\pgfplotsset{compat=newest}
\usetikzlibrary{plotmarks}
\usetikzlibrary{arrows.meta}
\usetikzlibrary{shapes.multipart}
\usetikzlibrary{shapes,arrows,calc,positioning}
\usepgfplotslibrary{patchplots}

\tikzstyle{block} = [draw, fill=white, rectangle, minimum height=3em, minimum width=4em]
\tikzstyle{sum} = [draw, fill=white, circle, node distance=1cm]

\newtheorem{theorem}{Theorem}
\newtheorem{property}{Property}

\newtheorem{remark}{Remark}

\newtheorem{lemma}{Lemma}

\DeclareMathOperator{\vect}{vec}
\newcommand{\Z}{\ensuremath{{\mathbb Z}}}
\newcommand{\N}{\ensuremath{{\mathbb N}}}
\newcommand{\R}{\ensuremath{{\mathbb R}}}
\newcommand{\Enc}{{\mathsf{Enc}}}
\newcommand{\Dec}{\mathsf{Dec}}
\newcommand{\Mult}{\mathsf{Mult}}
\newcommand{\Pack}{{\mathsf{Pack}}}
\newcommand{\Unpack}{{\mathsf{Unpack}}}
\newcommand{\Prod}{{\mathsf{Prod}}}

\newcommand{\Ceil}{\ensuremath{\left\lceil}}
\newcommand{\Flr}{\ensuremath{\right\rfloor}}
\newcommand{\col}{\ensuremath{\text{col}}}

\colorlet{Darkblue}{blue!70!black}

\begin{document}
	
\bstctlcite{IEEEexample:BSTcontrol}

\title{Encrypted Dynamic Control
	exploiting\\
	Limited Number of Multiplications
	and\\
	a Method
	using RLWE-based Cryptosystem}

\author{
Joowon Lee,~\IEEEmembership{Student Member,~IEEE,} Donggil Lee, Junsoo Kim,~\IEEEmembership{Member,~IEEE,} and Hyungbo Shim,~\IEEEmembership{Senior Member,~IEEE}
\thanks{This work was supported in part by the National Research Foundation of Korea(NRF) grant funded by the Korea government(MSIT) (No. RS-2022-00165417) and in part by the National Research Foundation of Korea(NRF) grant funded by the Korea government(MSIT) (No. RS-2024-00353032).}
\thanks{J.~Lee and H.~Shim are with ASRI, Department of Electrical and Computer Engineering, Seoul National University, Korea. D.~Lee is with the Department of Electrical Engineering, Incheon National University, Korea. J.~Kim is with the Department of Electrical and Information Engineering, Seoul National University of Science and Technology, Korea.}
}

\markboth{}%
{Author \MakeLowercase{\textit{et al.}}: Paper Title}


\maketitle

\begin{abstract}
In this paper, we present a method to encrypt dynamic controllers that can be implemented through most homomorphic encryption schemes, including somewhat, leveled fully, and fully homomorphic encryption. To this end, we represent the output of the given controller as a linear combination of a fixed number of previous inputs and outputs. As a result, the encrypted controller involves only a limited number of homomorphic multiplications on every encrypted data, assuming that the output is re-encrypted and transmitted back from the actuator. A guidance for parameter choice is also provided, ensuring that the encrypted controller achieves predefined performance for an infinite time horizon. Furthermore, we propose a customization of the method for Ring Learning With Errors (RLWE)-based cryptosystems, where a vector of messages can be encrypted into a single ciphertext and operated simultaneously, thus reducing computation and communication loads. Unlike previous results, the proposed customization does not require extra algorithms such as rotation, other than basic addition and multiplication. Simulation results demonstrate the effectiveness of the proposed method.
\end{abstract}


\section{Introduction}
\IEEEPARstart{W}{ith} the development of various attack methods targeting networked control systems \cite{secure,CPSsurvey},
confidentiality of such systems has gained importance to protect transmission data and the system model
from potential adversaries,
who attempt to gather private control data and generate more sophisticated attacks based on the collected information.
One of the approaches to protect significant data within the networked system is to use cryptography as in \cite{crypto}.
However,
not all cryptosystems enable direct computations on encrypted data,
thus putting the data in the middle of some operations on the network at risk of disclosure.

In this context, the notion of encrypted controller has been introduced, as in \cite{kogiso,Darup21CSM,ARC,Kim16necsys,Farokhi17},
where the controller operates directly over encrypted signals and parameters without decryption
through the use of homomorphic encryption (HE).
By doing so,
all private control data in the network
can be protected from the adversaries.
This also hinders the adversaries from
inferring some information about the plant model.
Thus, encryption of controllers leads to enhanced security against attacks that make use of model knowledge or disclosure resources, as classified in \cite{secure}.
Therefore, possible applications of encrypted control include a wide range of cyber-physical systems where the sensors measure private data or
the model information of the physical plant should be kept secure.

However,
implementation of dynamic encrypted controllers
is not straightforward
due to
the recursive nature of the state update in dynamic controllers,
while
the number of repeated homomorphic operations without decryption is limited in most cryptosystems.
Therefore,
several methods to encrypt dynamic controllers have been proposed in ways to avoid
recursive homomorphic operations being applied to the encrypted controller state.
Some of the early works
assumed that the whole state of the controller can be transmitted to the actuator and re-encrypted during each sampling period, as in \cite{kogiso}.
Here, re-encryption refers to
encryption of decrypted signals that are initially from the encrypted controller, and then sending them back to the controller, as depicted in Fig.~\ref{fig:enc sys} where $\bar{\mathbf{u}}(k)$ is the re-encrypted signal.
In case when the whole state is re-encrypted,
the re-encryption consumes heavy communication load as the state dimension grows.

Later on,
encrypted dynamic controllers that
transmit only the controller output, instead of the whole state,
have been presented
using only a limited number of repeated homomorphic operations.
For example, in \cite{Murguia20TAC},
the controller resets its state periodically in order to cease the ongoing recursive operations,
but this may result in performance degradation.
Meanwhile,
by applying re-encryption to the controller output,
it is shown in \cite{KimIFAC} that
the encrypted controller can operate non-recursively,
for controllers where
the output is a function of a finite number of previous inputs and outputs.

On the other hand,
research on encrypted dynamic controllers carrying recursive homomorphic operations
has been conducted,
utilizing techniques from cryptography or control theory.
In \cite{Kim16necsys},
the controller is encrypted by fully HE,
which enables any operation for an unlimited number of times
using the bootstrapping technique.
However, this technique has been regarded impractical for real-time operations in control systems due to its high computational complexity.
Subsequently,
in \cite{Kim22TAC},
a method to recursively update the encrypted state
without bootstrapping is proposed,
where only the controller output is re-encrypted.
In this case,
to implement the recursive multiplications,
a specific type of encryption scheme \cite{gsw} is utilized, which allows
the external product of encrypted data.

\subsection{Contribution and Outline}

In this paper,
we present
that encrypted linear dynamic controllers can be implemented
with any sort of encryption scheme
that allows a fixed number of addition and multiplication over encrypted messages,
while re-encrypting only the controller output.
Indeed, this applies to most
HE schemes
including somewhat, leveled fully, and fully HE.

To this end,
we first show that given a linear controller,
its output can be expressed
as a linear combination of a fixed number of previous inputs and outputs,
motivated by \cite{KimIFAC}.
Then,
by re-encrypting the controller output (instead of the whole state),
the operation of the controller becomes non-recursive
in the sense that homomorphic operations are executed only on newly encrypted data,
hence it can operate for an infinite time horizon.
The controller encrypted accordingly only involves quantization errors, which are generated by converting signals and control parameters into integer messages
so that they can be homomorphically encrypted.
We also provide a guidance for parameter design
ensuring that the
error between the encrypted and the given controller output to be arbitrarily small.

Furthermore,
we propose an encrypted controller design customized for Ring Learning With Errors (RLWE)-based cryptosystems,
which are widely used and accessible through libraries such as Microsoft SEAL \cite{seal}, OpenFHE \cite{OpenFHE},
and Lattigo \cite{lattigo}.
RLWE-based cryptosystems originate from LWE-based cryptosystems \cite{lwe}, but they are more efficient in terms of computation load and storage as they utilize the structure of polynomial rings.
Especially, a vector of messages can be encoded into a polynomial and homomorphically operated at once.
Our proposed design makes use of this property to reduce the communication load and the number of homomorphic operations performed by the controller.
Moreover, it does not involve algorithms or evaluation keys of RLWE-based cryptosystems other than homomorphic addition and multiplication,
unlike the previous result \cite{Teranishi} that requires rotation and key switching keys.
Numerical analysis on the computational burden and
simulation results demonstrate the efficiency and practicality of the proposed design.

The rest of the paper is organized as follows.
Section~\ref{sec:pre} provides preliminaries on HE and the problem formulation.
In Section~\ref{sec:swhe}, the design of encrypted dynamic controllers is presented.
In Section~\ref{sec:rlwe}, our customized design for RLWE-based cryptosystems is proposed.
Section~\ref{sec:discussion}
discusses the efficiency of the customized design and
Section~\ref{sec:simul} provides simulation results.
Finally, Section~\ref{sec:conclude} concludes the paper.

\subsection{Notation}

The sets of integers, positive integers, nonnegative integers, and real numbers are denoted by $\Z$, $\N$, $\Z_{\geq 0}$ and $\R$, respectively.
We define $\mathbb{Z}_N:=\left\lbrace z\in\mathbb{Z}|-N/2\leq z<N/2\right\rbrace$ for $N\in\N$.
Let $\lfloor \cdot \rfloor$, $\Ceil \cdot\Flr$, and $z\,\,\,\mathrm{mod}\,\,N:=z-\left\lfloor(z+N/2)/N\right\rfloor N$ for $z\in\Z$ denote the floor, rounding, and modulo operation, respectively, which are defined element-wisely for vectors and matrices.
A sequence of scalars, vectors, or matrices $a_1,\,\ldots,\,a_n$ is written as $\left\lbrace a_i \right\rbrace_{i=1}^n$,
and let $\col\left\lbrace a_i\right\rbrace_{i=1}^n:=\left[ a_1^\top,\,a_2^\top,\,\cdots,\,a_n^\top\right]^\top$.
 The Hadamard product of two column vectors
$a=\col\left\lbrace a_i\right\rbrace_{i=1}^n\in\R^n$ and $b=\col\left\lbrace b_i\right\rbrace_{i=1}^n\in\R^n$
is defined as $a\circ b:=\col\left\lbrace a_ib_i\right\rbrace_{i=1}^n$.
Let $\lVert \cdot\rVert$ denote the infinity norm of a matrix or a vector.
The vectorization of a matrix $A$ and the Kronecker product are written by $\vect(A)$ and $\otimes$, respectively.
We denote the zero column vector of length $n$, the $m\times n$ zero matrix, and the $n\times n$ identity matrix by $\mathbf{0}_n$, $\mathbf{0}_{m\times n}$, and $I_n$, respectively.

\section{Preliminaries \& Problem Formulation}\label{sec:pre}

\subsection{Required Homomorphic Properties}\label{subsec:swhe}
HE allows certain operations on plaintexts (unencrypted data), such as addition and multiplication, to be executed over ciphertexts (encrypted data).
Consider an HE scheme with encryption and decryption algorithm denoted by $\Enc:\mathcal{P}\rightarrow \mathcal{C}$ and
$\Dec:\mathcal{C}\rightarrow\mathcal{P}$, where
$\mathcal{P}$ and $\mathcal{C}$ are the space of plaintexts and ciphertexts, respectively.
Then, for an operation $*_\mathcal{P}$ over $\mathcal{P}$, there exists an operation $*_\mathcal{C}$ over $\mathcal{C}$ such that
\begin{align*}
	\Dec\left(\Enc(m_1)*_\mathcal{C}\Enc(m_2) \right)=m_1 *_\mathcal{P} m_2, \quad \forall m_1,\,m_2\in\mathcal{P}.
\end{align*}

Throughout the paper, we consider HE schemes which support both addition and multiplication over ciphertexts.
Though this excludes partial HE developed at the early stage,
it is known to be not secure against attacks using quantum computers \cite{quantum}.
This subsection introduces basic properties
that are
generally satisfied by quantum-resistant HE,
rather than dealing with specific schemes.
For a detailed introduction to LWE-based cryptosystems, please refer to \cite{bookchapter}.

The form of the plaintext space varies by cryptosystems,
but it
is generally based on a finite set of integers.
In this subsection, we
let the plaintext space be 
given as $\mathcal{P}=\Z_N$ with a parameter $N\in\N$.
Accordingly, the encryption and decryption algorithm satisfy
correctness, \textit{i.e.},
\begin{align}\label{eq:correct}
	\Dec\left(\Enc\left(m\right)\right)=m \,\,\,\mathrm{mod}\,\, N, \quad \forall m\in\mathcal{P}.
\end{align}

The additively homomorphic property refers that
there exists an operation $\oplus$ over ciphertexts such that
\begin{align}\label{eq:add}
	\Dec\left(\Enc\left(m_1\right)\oplus\Enc\left(m_2\right)\right)=m_1+ m_2 \,\,\,\mathrm{mod}\,\, N,
\end{align}
for any $m_1\in\mathcal{P}$ and $m_2\in\mathcal{P}$.
In addition, due to the multiplicatively homomorphic property, 
finite linear combinations of plaintexts can be computed over encrypted data, as described in the following property.

\begin{property}\upshape\label{prop:1}
	For given $N\in\N$ and $\bar{r}\in\N$, there exists an operation $\Prod_1$ over ciphertexts such that
	\begin{multline*}
		\Dec\left(\Prod_1\left(\left\lbrace \Enc(a_i)\right\rbrace_{i=1}^r, \left\lbrace \Enc(m_i)\right\rbrace_{i=1}^r\right)\right)
		\\=\sum_{i=1}^{r}a_im_i \,\,\,\mathrm{mod}\,\,N,
	\end{multline*}
	for any $a_i\in\Z_N$, $m_i\in\Z_N$, $i=1,\,2,\,\ldots,\,r$, and $r\leq \bar{r} $.\qed
\end{property}

In Property~\ref{prop:1},
once a plaintext is encrypted, it only undergoes a single homomorphic multiplication and at most $\bar{r}-1$ homomorphic additions.
Therefore, Property~\ref{prop:1} can be achieved by most HE schemes, even by somewhat HE
where the number of repeated homomorphic multiplications is very limited.

We define the element-wise encryption of integer vectors as
$\Enc_n(a):=\left\lbrace \Enc(a_i) \right\rbrace_{i=1}^n$
for any vector of plaintexts $a\in \Z_N^n$.
Likewise, the component-wise encryption of a matrix
$A=\{A_{ij}\}\in \Z_N^{m\times n}$
is defined as $\Enc_{m\times n}\left(A\right):=
\{\{\Enc(A_{ij})\}_{i=1}^{m}\}_{j=1}^{n}$,
as a collection of encrypted scalars.
For the decryption of collected ciphertexts, let us abuse notation and denote it as $\Dec(\cdot)$.

Homomorphic addition of element-wisely encrypted vectors or matrices is also defined element-wisely, satisfying \eqref{eq:add}.
Correspondingly, homomorphic multiplication between $\Enc_{m\times n}(A)$ and $\Enc_n(a)$
is defined as
\begin{subequations}
	\begin{equation}\label{eq:cprod}
		\Enc_{m\times n}\left(A\right)*\Enc_n(a):=\left\lbrace b_i\right\rbrace_{i=1}^n,
	\end{equation}
	where
	\begin{equation}
		b_i:=\Prod_1\left(\left\lbrace \Enc(A_{i,j})\right\rbrace_{j=1}^n, \Enc_n(a)\right),
	\end{equation}
\end{subequations}
if $n\leq \bar{r}$ for $\bar{r}$ given in Property~\ref{prop:1}.
Furthermore, we define the multiplication between a matrix of plaintexts $A\in\Z_N^{m\times n}$ and $\Enc_n(a)$ as
\begin{align}\label{eq:plainmtx}
	A\cdot\Enc_n\left( a\right) \!:=\!\left\lbrace A_{i,1}\Enc\left( a_1\right)\!\oplus\! \cdots\!\oplus\! A_{i,n}\Enc\left( a_n\right) \right\rbrace_{i=1}^m.
\end{align}

\subsection{RLWE-based Cryptosystem}\label{subsec:rlwe}

This subsection briefly describes RLWE-based cryptosystems and their key properties utilized in Section~\ref{sec:rlwe}.
RLWE-based cryptosystems make use of the structure of polynomial rings so that both plaintexts and ciphertexts consist of polynomials.
They include several well-known HE schemes such as BFV \cite{bfv}, BGV \cite{bgv}, and CKKS \cite{heaan},
which share common properties introduced in this subsection.

A polynomial ring $R_{p,N}:=\Z_N[X]/\langle X^p+1\rangle$
can be understood as the finite set of polynomials
with degree less than $p$ and coefficients in $\Z_N$.
Any integer polynomial can be mapped to a polynomial in this set
by taking the modular operation $\mathrm{mod}\left( X^p+1,N\right)$, where
the operation $\mathrm{mod}\,\,N$ is applied to each coefficient and $X^p$ is regarded as $-1$.

To encrypt (quantized) vector signals in control systems through a RLWE-based cryptosystem,
a method to encode an integer vector into a polynomial in $R_{p,N}$ called ``packing''\cite[Section 5.1.1]{bgv} can be utilized.
Based on the number theoretic transform,
the packing function $\Pack:\Z_N^p\rightarrow R_{p,N}$ and the unpacking function $\Unpack:R_{p,N}\rightarrow \Z_N^p$
satisfy the following properties;
\begin{equation*}
	\begin{aligned}
		&\mathsf{Unpack}\left(\mathsf{Pack}(u)\right)=u\,\,\,\mathrm{mod}\, N,\\
		&\mathsf{Unpack}\left(f(X)+g(X) \,\,\,\mathrm{mod}\,\left(X^p+1,N \right) \right)=u+v\,\,\,\mathrm{mod}\,N,\\
		&\mathsf{Unpack}\left(f(X)g(X)\,\,\,\mathrm{mod}\, \left(X^p+1,N \right)\right)=u\circ v\,\,\,\mathrm{mod}\, N,
	\end{aligned}
\end{equation*}
for any $u\in\Z_N^p$ and $v\in \Z_N^p$,
where $f(X)=\Pack(u)$ and $g(X)=\Pack(v)$.
See Appendix~A for more details and an example on the packing and unpacking functions.

In case when a RLWE-based cryptosystem is used,
we apply the packing function before every encryption, and similarly the unpacking function after every decryption.
This enables element-wise addition and multiplication between vectors to be computed over ciphertexts at once,
without having to encrypt each component of the vectors separately.

The basic settings and algorithms of RLWE-based cryptosystems are introduced below.
Refer to Appendix~B for the details of these algorithms in case of the BGV scheme.
\begin{itemize}
	\item \textit{Parameters} $\left(N,\,p,\,q\right)$:
	The plaintext space $\mathcal{P}$ is $R_{p,N}$ and
	the ciphertext space $\mathcal{C}$ is $R_{p,q}^2$ or $R_{p,q}^3$,
	where
	$q\gg N$,
	$N= 1\,\,\,\mathrm{mod}\,\,2p$,
	and $p$ is a power of $2$.
	\item \textit{Encryption and packing}:
	For $m\in\Z_N^p$, define
	$\Enc^\prime(m):=\Enc\left( \Pack\left(m \right) \right)\in R_{p,q}^2.$
	\item \textit{Decryption and unpacking}:
	For $\mathbf{c}\in R_{p,q}^2$ or $R_{p,q}^3$, define
	$\Dec^\prime(\mathbf{c}):=\Unpack\left( \Dec(\mathbf{c})\right)\in \Z_N^p.$
	\item \textit{Homomorphic addition} $\oplus:R_{p,q}^i\times R_{p,q}^i\to R_{p,q}^i$ for $i=2,\,3$.
	\item \textit{Homomorphic multiplication} $\Mult:R_{p,q}^2\times R_{p,q}^2\to R_{p,q}^3$.
\end{itemize}

Note that the homomorphic multiplication of RLWE-based cryptosystems increases the dimension of ciphertexts by one.
There exists an algorithm called relinearization \cite{bgv}
which reduces the dimension of ciphertexts to $2$ while preserving the message inside.
However, relinearization is not necessary for the proposed encrypted controller in this paper, so we handle ciphertexts of both length $2$ and $3$.

Given proper encryption parameters,
RLWE-based cryptosystems satisfy \eqref{eq:correct}, \eqref{eq:add}, and Property~\ref{prop:1},
since $a_i$ and $m_i$, $i=1,\,2,\,\ldots,\,r$, can be regarded as constant polynomials in $R_{p,N}$.
In addition,
the following property is satisfied,
where vectors,
instead of scalars, are encrypted.

\begin{property}\upshape\label{prop:2}
	For given $N\in\N$, $\bar{r}\in \N$, and $p\in\N$, there exists an operation $\Prod_2$ over ciphertexts such that
	\begin{multline}
		\Dec^\prime\left(\Prod_2\left(\left\lbrace \Enc^\prime\left(\mathbf{a}_i\right)\right\rbrace_{i=1}^r ,\left\lbrace \Enc^\prime\left(\mathbf{m}_i\right)\right\rbrace_{i=1}^r\right)\right)\\	=\sum_{i=1}^{r}\mathbf{a}_i\circ\mathbf{m}_i\,\,\,\mathrm{mod}\,\,N,
	\end{multline}\label{eq:prop2}
	for any $\mathbf{a}_i\in\Z_N^p$, $\mathbf{m}_i\in\Z_N^p$, $i=1,\,2,\,\ldots,\,r$, and $r\leq \bar{r}$.\qed
\end{property}

Property~\ref{prop:2} indicates that RLWE-based cryptosystems with packing support a single component-wise homomorphic multiplication of newly encrypted vectors followed by at most $\bar{r}-1$ homomorphic additions.
In contrast,
for cryptosystems having only Property~\ref{prop:1},
this can be achieved by repeating the operation $\Prod_1$ for $p$ times.

\subsection{Problem Formulation}
Consider a discrete-time plant written as
\begin{equation}\label{eq:plant}
	\begin{aligned}
		x_p(k+1)&=Ax_p(k)+Bu(k),\\
		y(k)&=Cx_p(k),
	\end{aligned}
\end{equation}
where $x_p(k)\in\R^{n_p}$, $u(k)\in\R^h$, and $y(k)\in\R^l$ is the state, input, and output of the plant, respectively.
Suppose that a discrete-time dynamic controller
has been designed as
\begin{equation}\label{eq:ctr}
	\begin{aligned}
		x(k+1)&=Fx(k)+Gy(k),\quad x(0)=x_0,\\
		u(k)&=Hx(k),
	\end{aligned}
\end{equation}
where $x(k)\in\mathbb{R}^n$ is the state,
so that the closed-loop system of \eqref{eq:plant} and \eqref{eq:ctr} is stable.
Throughout the paper, it is assumed that the controller \eqref{eq:ctr} is controllable and observable.

We aim to construct an encrypted controller from the given controller \eqref{eq:ctr} satisfying the followings:
\begin{itemize}
	\item
	All signals being transmitted between the plant and the controller are encrypted, and
	only the controller output is sent to the plant rather than the whole state.
	An additional communication link is installed to re-encrypt the controller output at the actuator and transmit it back to the controller, as shown in Fig.~\ref{fig:enc sys}.
	\item 
		Every entity on the network (the shaded area in Fig.~\ref{fig:enc sys}), including the honest-but-curious encrypted controller and external hackers,
        is not capable of decryption.  
	\item 
	It can be implemented through any cryptosystem satisfying Property~\ref{prop:1}, which corresponds to most HE schemes, including somewhat, leveled fully, and fully HE.
	\item
	The proposed design guarantees that
	the error between the output of the original controller and that of the encrypted controller can be made arbitrarily small,
	by adjusting parameters for quantization.
\end{itemize}

\section{Encrypted Controller Design}\label{sec:swhe}

In this section, we present a design method of encrypted controllers
which can be realized through any cryptosystem
that supports finite homomorphic linear combinations over newly encrypted data.
In order to utilize such cryptosystems, the encrypted controller is designed to perform a limited number of homomorphic operations at each time step.
Therefore, only Property~\ref{prop:1} and the basic homomorphic properties stated in Section~\ref{subsec:swhe} are used throughout this section.

To this end,
we transform
the controller \eqref{eq:ctr} first so that the output is represented using a fixed number of previous inputs and outputs,
by feeding back the output itself.
Next, based on this transformed controller, we design the encrypted controller that achieves the desired control performance by proper choice of parameters.

We define a new state for the given controller \eqref{eq:ctr} which consists of the inputs and outputs during the past $n$ steps, by
\begin{multline*}
	z(k):=\left[
	y(k-1)^\top,\,\ldots,\,y(k-n)^\top,\right.\\
	\left.u(k-1)^\top,\,\ldots,\,u(k-n)^\top\right]^\top\in\R^{\bar{n}},
\end{multline*}
where $\bar{n}:=n(h+l)$.
Since $z(k)$ has an increased dimension compared to the original state $x(k)$,
the following lemma is provided to ensure the existence of a mapping from $z(k)$ to $x(k)$,
before expressing the controller \eqref{eq:ctr} with $z(k)$.

\begin{lemma}\upshape\label{lem:M}
	If the controller \eqref{eq:ctr} is
	controllable and observable,
	then there exist $M\in\R^{n\times\bar{n}}$ and
	$z_0\in\R^{\bar{n}}$ such that $x(k)=Mz(k)$ for all $k\in\Z_{\geq 0}$, with $z(0)=z_0$.\qed
\end{lemma}

\renewcommand\qedsymbol{$\blacksquare$}
\begin{proof}
	By the observability of $(F,H)$, there exists a matrix $R\in\R^{n\times h}$ such that $\bar{F}:=F-RH$ is nilpotent.
	Then,
	it follows that
	\begin{align*}
		x(k+1)=\bar{F}x(k)+Gy(k)+Ru(k),\quad u(k)=Hx(k).
	\end{align*}
	Since $\bar{F}^n=\mathbf{0}_{n\times n}$, the state $x(k)$ can be computed as
	\begin{align}\label{eq:3a-12}
		x(k)=\sum_{i=1}^{n}\bar{F}^{i-1}\left( G y(k-i)+R u(k-i)\right)=:Mz(k),
	\end{align}
	for all $k\geq n$.
	Suppose that $x(-n)=\mathbf{0}_n$, then there exists an input sequence $\left\lbrace y(k)\right\rbrace_{k=-n}^{-1} $ which satisfies $x(0)=x_0$
	for any $x_0\in\R^n$
	by the controllability.
	Accordingly, the output sequence $\left\lbrace u(k)\right\rbrace_{k=-n}^{-1} $ 
	and then $z(k)$ for $k=0,\,1,\,\ldots,\,n-1$ are determined, satisfying \eqref{eq:3a-12} for all $k\in\Z_{\geq 0}$.
	Let $z_0$ be defined as the determined $z(0)$, thus concluding the proof.
\end{proof}
\renewcommand\qedsymbol{$\square$}

From the initial value $z_0$ given by Lemma~\ref{lem:M}, we define $u(k)$ and $y(k)$ for $k=-1,\,-2,\,\ldots,\,-n$ virtually as
\begin{align}\label{eq:minus}
	z_0=:\left[y(-1)^\top\!,\,\ldots,\,y(-n)^\top\!,\,u(-1)^\top\!,\,\ldots,\,u(-n)^\top\right]^\top\!.
\end{align}
Then, using the matrix $M$ from Lemma~\ref{lem:M},
the controller \eqref{eq:ctr} is transformed into
\begin{subequations}\label{eq:zdyn}
	\begin{align}			z(k+1)&=\mathcal{F}z(k)+\mathcal{G}y(k)+\mathcal{R}u(k),\quad z(0)=z_0,\label{eq:z}\\
		u(k)&=\mathcal{H}z(k),\label{eq:uz}
	\end{align}
\end{subequations}
where    
\begin{align*}
	\mathcal{F}&:=\left[\begin{array}{lll|lll}
		\multicolumn{2}{c}{\mathbf{0}_{l\times (n-1)l}} & \mathbf{0}_{l\times l} &  &  & \\
		& & &  \multicolumn{3}{c}{\mathbf{0}_{nl\times nh}}\\
		\multicolumn{2}{c}{\smash{\raisebox{.5\normalbaselineskip}{$I_{(n-1)l}$}}} &
		{\smash{\raisebox{.5\normalbaselineskip}{$\mathbf{0}_{(n-1)l\times l}$}}} & & &\\
		\hline
		& & & \multicolumn{2}{c}{\mathbf{0}_{h\times (n-1)h}} & \mathbf{0}_{h\times h}\\
		\multicolumn{3}{c}{\mathbf{0}_{nl\times nh}} \vline  & & &\\
		& & &
		\multicolumn{2}{c}{\smash{\raisebox{.5\normalbaselineskip}{$I_{(n-1)h}$}}} &
		{\smash{\raisebox{.5\normalbaselineskip}{$\mathbf{0}_{(n-1)h\times h}$}}}
	\end{array} \right], \\
	\mathcal{G}&:=\left[ \begin{array}{c}
		I_l\\
		\mathbf{0}_{(n-1)l\times l}\\
		\hline \\[-\normalbaselineskip]
		\\
		{\smash{\raisebox{.5\normalbaselineskip}{$\mathbf{0}_{nl\times h}$}}}
	\end{array}\right],\,\,
	\mathcal{R}:=\left[ \begin{array}{c}
		\\
		{\smash{\raisebox{.5\normalbaselineskip}{$\mathbf{0}_{nl\times h}$}}}\\
		\hline \\[-\normalbaselineskip]
		I_l\\
		\mathbf{0}_{(n-1)h\times h}
	\end{array}\right],\,\,
	\mathcal{H}:=HM.
\end{align*}
It can be observed that \eqref{eq:z} represents the update of $z(k)$ by definition,
where $y(k)$ and $u(k)$ are placed at the top of the first $nl$ and the last $nh$ elements of $z(k)$ by $\mathcal{G}$ and $\mathcal{R}$, respectively.
Note that $\mathcal{F},\,\mathcal{G}$, and $\mathcal{R}$ are integer matrices consisting only of zeros and ones.

Based on \eqref{eq:zdyn},
we construct the encrypted controller as follows.
First, the control parameter $\mathcal{H}$
is quantized with a parameter $1/s\geq 1$ and then encrypted as
\begin{align*}
	\mathbf{H}:=\Enc_{h\times \bar{n}}\left( \Ceil \frac{\mathcal{H}}{s}\Flr\right).
\end{align*}
Since the structures of $\mathcal{F}$, $\mathcal{G}$, and $\mathcal{R}$ are universal by construction
and do not depend on the model of the given controller,
we leave them unencrypted.

The sensor and the actuator encrypt the plant output and the input with a parameter $1/L>0$ for quantization, as
\begin{align}\label{eq:boldy}
	\mathbf{y}(k)=\mathsf{Enc}_l\left(\left\lceil \frac{y(k)}{L}\right\rfloor\right),\,\,
	\mathbf{u}(k)=\Enc_h\left(\Ceil \frac{u(k)}{L}\Flr\right),
\end{align}
respectively.
The encrypted controller receives $\mathbf{y}(k)$ and $\mathbf{u}(k)$, then returns $\bar{\mathbf{u}}(k)$ as follows;
\begin{subequations}\label{eq:3}
	\begin{align}
		\mathbf{z}(k+1)&=\mathcal{F}\cdot\mathbf{z}(k)\oplus\mathcal{G}\cdot\mathbf{y}(k)\oplus\mathcal{R}\cdot\mathbf{u}(k),\label{eq:3a}\\
		\bar{\mathbf{u}}(k)&=\mathbf{H}* \mathbf{z}(k),\label{eq:3b}\\ \mathbf{z}(0)&=\Enc_{\bar{n}}\left( \Ceil \frac{z_0}{L}\Flr\right),\notag
	\end{align}
\end{subequations}
where the operators $\oplus$, $\cdot$, and $\ast$ are defined in \eqref{eq:add}, \eqref{eq:plainmtx}, and \eqref{eq:cprod}, respectively.
By the definitions of $\mathcal{F}$, $\mathcal{G}$, and $\mathcal{R}$, it can be observed from \eqref{eq:3a} that the state $\mathbf{z}(k)$ acts as a container storing $n$ pairs of encrypted inputs and outputs, as
\begin{align*}
	\mathbf{z}(k)=\begin{bmatrix}
		\left\lbrace\mathbf{y}(k-i)\right\rbrace_{i=1}^n \\
		\left\lbrace\mathbf{u}(k-i)\right\rbrace_{i=1}^n
	\end{bmatrix}.
\end{align*}
Thus, the controller \eqref{eq:3} does not exhibit recursive
homomorphic operations when updating the state $\mathbf{z}(k)$.

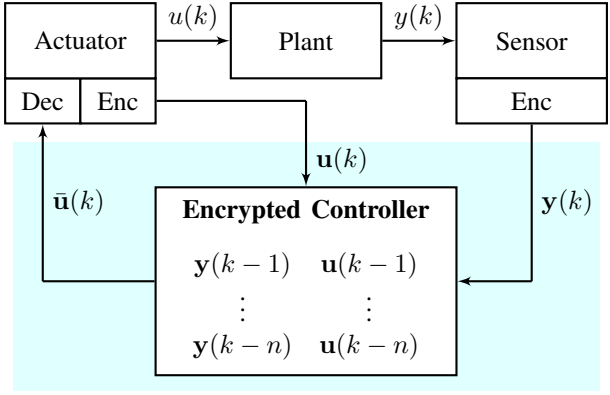
\begin{figure}[t!]
	\centering
 \scalebox{0.95}{
	\begin{tikzpicture}[node distance=2cm,>=latex']
		
		\node[rectangle, minimum height = 3.3cm, minimum width = 7.8cm, very thick, fill = LightCyan] (network) at (3.5,-2.2) {};
		
		\node[block, name = decryptor, thick, minimum height= 0.6cm, minimum width = 1cm]{Dec};
		\node[block, name = encryptu, thick, minimum height = 0.6cm, right of = decryptor, minimum width = 1cm, node distance = 1cm]{Enc};
		\node[coordinate, name = act, right of = decryptor, node distance = 0.5cm]{};
		\node[block, above of = act, name = actuator, minimum width = 2cm, thick, minimum height = 1cm, node distance = 0.8cm]{\large Actuator};
		\node[block, right of = actuator, name = plant, thick, minimum height = 1cm, minimum width = 2cm, node distance = 3cm]{\large Plant};
		\node[block, right of = plant, name = sensor, thick, minimum height = 1cm, minimum width = 2cm, node distance = 3cm]{\large Sensor};
		\node[block, below of = sensor, name = encrypty, thick, minimum height = 0.6cm, node distance = 0.8cm, minimum width = 2cm]{Enc};
		\node[block, below of = plant, name = controller, thick, node distance = 3.2cm, align=center, minimum width = 4cm, minimum height = 2.5cm]{
			\textbf{Encrypted Controller}\\ \\
			$\begin{matrix}
				\mathbf{y}(k-1)\\ \vdots \\ \mathbf{y}(k-n)
			\end{matrix}\quad \begin{matrix}
			\mathbf{u}(k-1)\\ \vdots \\ \mathbf{u}(k-n)
		\end{matrix}$
		};
		\node[coordinate, below of = encrypty, name = boldy, node distance = 2.4cm]{};
		\node[coordinate, below of = decryptor, name = boldu, node distance = 2.4cm]{};
		\node[coordinate, below of = plant, name = reenc, node distance = 0.8cm]{};
		
		\draw[->, thick] (actuator) -- (plant) node[pos = 0.5, above]{$u(k)$};
		\draw[->, thick] (plant) -- (sensor) node[pos = 0.5, above]{$y(k)$};
		\draw[-, thick] (encrypty) -- (boldy) node[pos = 0.5, right]{$\mathbf{y}(k)$};
		\draw[->, thick] (boldy) -- (controller);
		\draw[->, thick] (boldu) -- (decryptor) node[pos = 0.5, right]{$\bar{\mathbf{u}}(k)$};
		\draw[-, thick] (controller) -- (boldu);
		\draw[-, thick] (encryptu) -- (reenc);
		\draw[->, thick] (reenc) -- (controller) node[pos = 0.7, right]{$\mathbf{u}(k)$};
	\end{tikzpicture}
 }
	\caption{System configuration with the encrypted controller \eqref{eq:3}-\eqref{eq:actu} exploiting only a limited number of homomorphic multiplications.
 The shaded area represents the networked part of the system.}
	\label{fig:enc sys}
\end{figure}

In \eqref{eq:3a}, the scale of $\mathbf{z}(k)$ is maintained to be $1/L$
due to its initial value and the matrices $\mathcal{F}$, $\mathcal{G}$, and $\mathcal{R}$ consisting only of integer components.
On the other hand, the output $\bar{\mathbf{u}}(k)$ in \eqref{eq:3b} is of scale $1/(Ls)$
because $\mathcal{H}$ is scaled by $1/s$.
Therefore, the actuator decrypts $\bar{\mathbf{u}}(k)$, re-scales it, and then returns the plant input during the re-encryption process, as
\begin{align}\label{eq:actu}
	u(k)=\Dec\left(\bar{\mathbf{u}}(k)\right)\cdot Ls.
\end{align}
The overall encrypted control system is depicted in Fig.~\ref{fig:enc sys}.

Now we analyze the performance of the encrypted controller \eqref{eq:3}.
To begin with, consider a perturbed controller written by
\begin{equation}\label{eq:xe}
	\begin{aligned}
		x(k+1)&=Fx(k)+Gy(k)+e_x(k),\\
		u(k)&=Hx(k)+e_u(k),\quad
		x(0)=x_0+e_0,
	\end{aligned}
\end{equation}
where $e_x(k)\in\R^n$, $e_u(k)\in\R^h$, and $e_0(k)\in\R^n$ are perturbations added to the original controller \eqref{eq:ctr}.
The virtual values of $u(k)$ and $y(k)$ for $k=-1,\,-2,\,\ldots,\,-n$ are again defined as \eqref{eq:minus}, without the perturbations.

We claim that the messages inside \eqref{eq:3} obey the same dynamics as \eqref{eq:xe}, regarding the quantization errors as perturbations.
It can be assured by showing that the signals quantized in \eqref{eq:boldy},
the message inside the initial state $\mathbf{z}(0)$,
and the outcome of the operations in \eqref{eq:xe} belong to the plaintext space $\Z_N$ for the whole time.
If this was not the case, they would be modified by the modular operation during the decryption in \eqref{eq:actu} (recall \eqref{eq:correct}).
To this end, we first choose the modulus $N$ to satisfy
\begin{equation}\label{eq:N}
	\frac{1}{L}\max\left\lbrace \frac{\lVert u(k)\rVert}{s},\,\lVert y(k)\rVert,\,\lVert z_0\rVert\right\rbrace+\frac{1}{2}<\frac{N}{2}.
\end{equation}
For now, \eqref{eq:N} can only be satisfied for a finite time horizon, since we have not shown the boundedness of $u(k)$ and $y(k)$ in \eqref{eq:xe} yet.

The following lemma shows that the encrypted controller \eqref{eq:3}-\eqref{eq:actu} is equivalent to \eqref{eq:xe} with bounded perturbations,
while the condition \eqref{eq:N} is satisfied.

\begin{lemma}\upshape\label{lem:perturb}
	Given $T\in\Z_{\geq 0}$, suppose that
	the perturbed controller \eqref{eq:xe} satisfies
	\eqref{eq:N} for all $k\in\left\lbrace 0,\,1,\,\ldots,\,T \right\rbrace$.
 	Then,
	the encrypted controller \eqref{eq:3}-\eqref{eq:actu}
	generates the same control input sequence $\left\lbrace u(k) \right\rbrace_{k=0}^T $
	as \eqref{eq:xe},
	with some $\left\lbrace  e_0,\, e_x(k),\, e_u(k)\right\rbrace$ satisfying
	\begin{multline}\label{eq:eubound}
		\lVert e_0\rVert \leq \frac{L}{2}\lVert M\rVert,\quad
		\lVert e_x(k)\rVert \leq \frac{L}{2}\lVert M\rVert,\quad\text{and}\\
		\lVert e_u(k)\rVert \leq \frac{s\bar{n}}{2}\left(\left\lVert\col\left\lbrace \begin{bmatrix}
			y(k-i) \\u(k-i)
		\end{bmatrix}\right\rbrace_{i=1}^n\right\rVert +\frac{L}{2} \right).
	\end{multline}\qed
\end{lemma}

\renewcommand\qedsymbol{$\blacksquare$}
\begin{proof}
	We first specify the perturbations in \eqref{eq:xe} and show that \eqref{eq:eubound} holds.
	To this end, consider a controller written as
	\begin{subequations}\label{eq:ze}
	\begin{align}
		z(0)&=L\Ceil \frac{z_0}{L}\Flr=:z_0+e_{0,z},\\
		z(k+1)&=\mathcal{F}z(k)+L\cdot\mathcal{G}\Ceil \frac{y(k)}{L}\Flr+ L\cdot\mathcal{R}\Ceil \frac{u(k)}{L}\Flr\nonumber\\
    &=:\mathcal{F}z(k)+\mathcal{G}y(k)+\mathcal{R}u(k)+e_z(k),\label{eq:zez}\\
		u(k)&=s\Ceil\frac{\mathcal{H}}{s}\Flr z(k)=:\mathcal{H}z(k)+e_{u,z}(k)\label{eq:zeu},
	\end{align}
	\end{subequations}
	which is in the form of \eqref{eq:zdyn} with perturbations $e_{0,z}\in\R^{\bar{n}}$, $e_z(k)\in\R^{\bar{n}}$, and $e_{u,z}(k)\in\R^h$.
	The perturbations can be expressed explicitly as
	\begin{align}\label{eq:ez}
			e_z(k)&=L\begin{bmatrix}
				\Ceil \frac{y(k)}{L}\Flr - \frac{y(k)}{L}\\
				\mathbf{0}_{(n-1)l}\\
				\Ceil \frac{u(k)}{L}\Flr - \frac{u(k)}{L}\\
				\mathbf{0}_{(n-1)h}\\
			\end{bmatrix},\\
			e_{u,z}(k)&=s\left( \Ceil \frac{\mathcal{H}}{s}\Flr-\frac{\mathcal{H}}{s}\right) z(k),\,\, e_{0,z}=L\left( \Ceil \frac{z_0}{L}\Flr -\frac{z_0}{L}\right),\nonumber
	\end{align}
	and hence are bounded as
	\begin{align*}
		\lVert e_z(k)\rVert\leq\frac{L}{2},\,\,\,
		\lVert e_{0,z}\rVert\leq\frac{L}{2},\,\,\,
		\text{and}\,\,\,
		\lVert e_{u,z}(k)\rVert \leq \frac{s\bar{n}}{2}\lVert z(k)\rVert.
	\end{align*}
	By multiplying $M$ to \eqref{eq:zez} and using the relation $x(k)=Mz(k)$, the controller \eqref{eq:ze} is transformed to \eqref{eq:xe} since $M\left( \mathcal{F}+\mathcal{R}\mathcal{H}\right)=FM$ and $M\mathcal{G}=G$.
	The perturbations are also transformed to $e_x(k)=Me_z(k)$, $e_0=Me_{0,z}$, and $e_u(k)=e_{u,z}(k)$.
	It is clear from \eqref{eq:ze} that
	\begin{align}\label{eq:zezz}
		z(k)=\begin{bmatrix}
			\col\left\lbrace \Ceil y(k-i)/L\Flr\right\rbrace_{i=1}^n\\
			\col \left\lbrace \Ceil u(k-i)/L\Flr\right\rbrace_{i=1}^n  
		\end{bmatrix}
	\end{align}
	for all $k\in\Z_{\geq 0}$, and therefore \eqref{eq:eubound} follows.
	
	For the rest of the proof, we denote $u(k)$ and $y(k)$ of \eqref{eq:ze} by $\tilde{u}(k)$ and $\tilde{y}(k)$, respectively,
	in order to differentiate them from $u(k)$ and $y(k)$ of \eqref{eq:boldy} and \eqref{eq:actu}.
	Now we prove that \eqref{eq:3} is equivalent to \eqref{eq:ze} as long as $\tilde{u}(k)$ and $\tilde{y}(k)$ satisfy \eqref{eq:N} for $k=0,\,1,\,\ldots,\,T$.
	Using the homomorphic properties \eqref{eq:add} and \eqref{eq:plainmtx},
	it is derived from \eqref{eq:boldy} and \eqref{eq:3} that
	\begin{align}\label{eq:boldz}
		\mathbf{z}(k)= \begin{bmatrix}
		\col\left\lbrace\Enc_l\left( \Ceil y(k-i)/L\Flr\right) \right\rbrace_{i=1}^n\\
		\col\left\lbrace\Enc_h\left(\Ceil u(k-i)/L\Flr\right)\right\rbrace_{i=1}^n
		\end{bmatrix},\quad \forall k\in\Z_{\geq 0}.
	\end{align}
	In addition, one can obtain from \eqref{eq:cprod} and Property~\ref{prop:1} that
	\begin{align}
		\Dec\left( \bar{\mathbf{u}}(k)\right)&= \left\lbrace \!\Dec \!\left( \Prod_1\!\left( \!\left\lbrace \Enc\left( \Ceil \mathcal{H}_{ij}/s\Flr\right) \right\rbrace_{j=1}^{\bar{n}}, \mathbf{z}(k)\right)\! \right)\! \right\rbrace_{i=1}^h\nonumber\\
		&=\Ceil \mathcal{H}/s\Flr \Dec\left( \mathbf{z}(k)\right) \,\,\,\mathrm{mod}\,\,N,\label{eq:boldubar}
	\end{align}
	where $\mathcal{H}_{ij}$ denotes the $(i,j)$-th component of the matrix $\mathcal{H}$.
	
	We show by induction that $z(k)=L\Dec\left( \mathbf{z}(k)\right)$, $u(k)=\tilde{u}(k)$, and $y(k)=\tilde{y}(k)$ for $k=0,\,1,\,\ldots,\,T$.
	Since \eqref{eq:N} implies that
	both $ \Ceil z_0/L\Flr$ and $\Ceil \tilde{u}(k)/(Ls)\Flr$ are in $\Z_N$, the decryption of \eqref{eq:boldz} leads to $z(0)/L$ without being altered by the modulo operation.
	Hence, it follows that
	\begin{align}\label{eq:proof0}
	u(0)=Ls\left( \Ceil \frac{\mathcal{H}}{s}\Flr \frac{z(0)}{L} \,\,\,\mathrm{mod}\,\,N\right) =\tilde{u}(0)
	\end{align}
	from \eqref{eq:actu} and \eqref{eq:boldubar},
	and thus $y(0)=\tilde{y}(0)$ under the same plant \eqref{eq:plant}.
	Suppose that the induction hypothesis holds for all $k=0,\,1,\,\ldots,\,\tau$, where $\tau$ is smaller than $T$.
	Then, since $u(\tau)$ and $y(\tau)$ satisfy \eqref{eq:N},
	the decryption of \eqref{eq:boldz} equals to \eqref{eq:zezz} at $k=\tau+1$.
	Therefore, by computing $u(\tau+1)$ from $z(\tau+1)$ analogously to \eqref{eq:proof0}, the induction concludes.
\end{proof}
\renewcommand\qedsymbol{$\square$}

So far, both the upper bound of $\lVert e_u(k)\rVert$ provided by \eqref{eq:eubound} and the lower bound of $N$ assumed by \eqref{eq:N} depend on the input and output
of the perturbed controller \eqref{eq:xe}.
However, using the closed-loop stability of \eqref{eq:plant} and \eqref{eq:ctr},
not only the constant bounds on the perturbations and the parameter $N$
but also the performance error between the given controller \eqref{eq:ctr} and its encryption \eqref{eq:3}
can be derived deterministically for an infinite time horizon.

To state the result, we consider two closed-loop systems; one consists of the plant \eqref{eq:plant} and the given controller \eqref{eq:ctr}, and the other consists of the plant \eqref{eq:plant} and the encrypted controller \eqref{eq:3}.
For clarity, let us denote the plant input and the output of the former closed-loop system as $u^\prime(k)$ and $y^\prime(k)$, respectively.
Since the original closed-loop system is stable, both $\left\lVert u^\prime(k)\right\rVert$ and $\left\lVert y^\prime(k)\right\rVert$ are bounded by some $S>0$ for all $k\in\Z_{\geq 0}$.

The following theorem provides an upper bound of the performance error,
in terms of the difference between $u(k)$ and $u^\prime(k)$,
which can be made arbitrarily small by adjusting the quantization parameters $L$ and $s$.

\begin{theorem}\upshape\label{thm}
	There exists\footnote{See (\ref{eq:epsilon}) in the proof for an explicit form of $\left\lbrace \epsilon_0,\,\epsilon_1,\,\epsilon_2,\,\epsilon_3\right\rbrace $.}
	a set of positive numbers
	$\left\lbrace \epsilon_0,\,\epsilon_1,\,\epsilon_2,\,\epsilon_3\right\rbrace $
	such that
	the encrypted controller \eqref{eq:3}-\eqref{eq:actu} guarantees
	\begin{align}\label{eq:thm}
		\left\lVert \begin{bmatrix}
			u(k) - u^\prime(k)\\
			y(k) - y^\prime(k)
		\end{bmatrix}\right\rVert
		\leq \frac{\epsilon_1L+\epsilon_2Ls+\epsilon_3s}{1-\epsilon_0s}=:\epsilon(L,s)
	\end{align}
	for all $k\in\Z_{\geq 0}$, provided that $N$, $L$, and $s$ satisfy
	\begin{align}\label{eq:Nbound}
		\frac{1}{L}\max\left\lbrace \frac{\epsilon(L,s)+S}{s},\,\lVert z_0\rVert\right\rbrace+\frac{1}{2}<\frac{N}{2}
	\end{align}
	and $1/s>\epsilon_0$.\qed
\end{theorem}

\renewcommand\qedsymbol{$\blacksquare$}
\begin{proof}
Under the closed-loop stability of the plant \eqref{eq:plant} and the perturbed controller \eqref{eq:xe},
	we show that \eqref{eq:N} and \eqref{eq:thm} hold for all $k\in\Z_{\geq 0}$
	if the perturbations satisfy \eqref{eq:eubound}.
	The closed-loop system of \eqref{eq:plant} and \eqref{eq:xe} is written by
	\begin{equation}\label{eq:cl}
		\begin{aligned}
			\mathsf{x}(k+1)\!&=\!\begin{bmatrix}
				A & BH\\
				GC & F
			\end{bmatrix}\!\!\begin{bmatrix}
				x_p(k) \\ x(k)
			\end{bmatrix}
			\!\!+\!\!\begin{bmatrix}
				B & \mathbf{0}_{n_p\times n}\\
				\mathbf{0}_{n\times h} & I_n
			\end{bmatrix}\! e(k)\\
			&=: \mathsf{A}\mathsf{x}(k) + \mathsf{B}e(k),\\
			\begin{bmatrix}
				y(k)\\u(k)
			\end{bmatrix}\!&=\!\begin{bmatrix}
				C & \mathbf{0}_{l\times n}\\
				\mathbf{0}_{h\times n_p} & H
			\end{bmatrix}\!\mathsf{x}(k)\!+\!\begin{bmatrix}
				\mathbf{0}_{l\times h} & \mathbf{0}_{l\times n}\\
				I_h & \mathbf{0}_{h\times n}
			\end{bmatrix}\! e(k)
			\\
			&=: \mathsf{C}\mathsf{x}(k) + \mathsf{D}e(k),
		\end{aligned}
	\end{equation}
	where the stacked perturbation $e(k):=\col\left\lbrace e_u(k),\,e_x(k)\right\rbrace $ is regarded as the external input,
	and
	the initial state is
	$\mathsf{x}(0)=\col\left\lbrace x_p(0),\,x_0+e_0\right\rbrace=: \mathsf{x}_0+\col\left\lbrace \mathbf{0}_{n_p},\,e_0\right\rbrace $.
	Since the matrix $\mathsf{A}$ is Schur stable, there exist $\alpha \geq 0$ and $\gamma\in[ 0,\,1) $ such that
	$\lVert \mathsf{A}^k\rVert\leq \alpha\gamma^k$ for all $k\in \Z_{\geq 0}$.
	Then, for all $k\in\Z_{\geq 0}$,
	\begin{equation}\label{eq:cl bound}
		\begin{aligned}
			\lVert \mathsf{x}(k+1)\rVert &\leq \alpha \lVert \mathsf{x}(0)\rVert+ \frac{\alpha\lVert \mathsf{B}\rVert}{1-\gamma}\max_{i\in [0,k]} \left\lbrace\lVert e(i) \rVert\right\rbrace,\\
			\left\lVert \begin{bmatrix}
				y(k)\\ u(k)
			\end{bmatrix}\right\rVert
			&\leq \alpha\lVert\mathsf{C}\rVert\lVert \mathsf{x}(0)\rVert+ \beta  \max_{i\in [0,k]}\left\lbrace\lVert e(i) \rVert\right\rbrace,
	\end{aligned}\end{equation}
	where $\beta:=1+\alpha\lVert\mathsf{C}\rVert\lVert\mathsf{B}\rVert/(1-\gamma)$, since $\lVert \mathsf{D}\rVert = 1$.
	
	Assuming $1/s>\bar{n}\beta/2$,
	we show by induction that
	\begin{align}\label{eq:ebound}
		\left\lVert e(k)\right\rVert \leq \Delta:=\max\left\lbrace \frac{L}{2}\lVert M\rVert,\,\frac{s\bar{n}}{2}\left(\lVert z_0\rVert+\frac{L}{2} \right),\,\delta \right\rbrace,
	\end{align}
	for all $k\in\Z_{\geq 0}$, where
	\begin{align*}
		\delta&:=
		\left( 1-\frac{s\bar{n}\beta}{2}\right) ^{-1}
		\frac{s\bar{n}}{2}\left( \alpha\lVert\mathsf{C}\rVert\lVert\mathsf{x}_0\rVert+\frac{L}{2}\alpha\lVert\mathsf{C}\rVert\lVert M\rVert+\frac{L}{2}\right).
	\end{align*}
	It is clear that \eqref{eq:ebound} holds at $k=0$ by \eqref{eq:eubound}.
	Suppose that \eqref{eq:ebound} holds for $k=0,\,1,\,\ldots,\,\tau$ with some $\tau\in\N$.
	By \eqref{eq:cl bound},
	it is derived that for $k=0,\,1,\,\ldots,\,\tau$,
	\begin{align*}
		\left\lVert \begin{bmatrix}
			y(k)\\ u(k)
		\end{bmatrix}\right\rVert \leq \alpha\lVert\mathsf{C}\rVert\left(\lVert \mathsf{x}_0\rVert + \frac{L}{2}\lVert M\rVert\right)+\beta \Delta=:\mathcal{U}(\Delta),
	\end{align*}	 
	since $\lVert \mathsf{x}(0)\rVert \leq \lVert \mathsf{x}_0\rVert + \lVert e_0\rVert$.
	Thus, we obtain
	\begin{align*}
		\frac{s\bar{n}}{2}\left(\left\lVert \begin{bmatrix}
			y(k)\\
			u(k)
		\end{bmatrix}\right\rVert+\frac{L}{2}\right)\leq \frac{s\bar{n}}{2}\left(\mathcal{U}(\Delta)+\frac{L}{2}\right)
		\leq \Delta
	\end{align*}
	for $k=\tau-n+1,\,\ldots,\,\tau-1,\,\tau$,
	where the last inequality results from the definition of $\delta$.
	This leads to $\lVert e_u(\tau+1)\rVert \leq \Delta$ by \eqref{eq:eubound}
	and proves \eqref{eq:ebound}.
	
	Next, consider the error dynamics defined by subtracting the closed-loop system of \eqref{eq:plant} and \eqref{eq:ctr} from \eqref{eq:cl}.
	It has the initial state $\col\left\lbrace \mathbf{0}_{n_p},\,e_0\right\rbrace $
	and the output bounded as
	\begin{align*}
		\left\lVert \begin{bmatrix}
			y(k)-y^\prime(k)\\ u(k)-u^\prime(k)
		\end{bmatrix}\right\rVert \leq \frac{L}{2}\alpha \lVert\mathsf{C}\rVert\lVert M\rVert+\beta \Delta=:\hat{\mathcal{U}}(\Delta).
	\end{align*}
	Then, there exists $\left\lbrace \epsilon_0,\,\epsilon_1,\,\epsilon_2,\,\epsilon_3\right\rbrace $ such that $\epsilon(L,s)\geq \hat{\mathcal{U}}(\Delta)$ and $\epsilon_0\geq \bar{n}\beta/2$, such as
	\begin{align}\label{eq:epsilon}
		\begin{bmatrix}
			\epsilon_0\\ \epsilon_1\\ \epsilon_2\\ \epsilon_3
		\end{bmatrix}=\frac{1}{2}\begin{bmatrix}
			\bar{n}\beta\\
			\lVert M\rVert \left( \alpha\lVert \mathsf{C}\rVert+\beta\right) \\
			\bar{n}\beta/2\\
			\bar{n}\beta\left( \alpha\lVert\mathsf{C}\rVert\lVert\mathsf{x}_0\rVert +\lVert z_0\rVert \right)
		\end{bmatrix}.
	\end{align}
	Therefore, the perturbed controller satisfies \eqref{eq:thm}, and hence 
	\eqref{eq:N} holds for all $k\in\Z_{\geq 0}$ by \eqref{eq:Nbound}.
	
	Since the system \eqref{eq:cl} satisfies \eqref{eq:N} for all $T\in\Z_{\geq 0}$,
	the encrypted controller \eqref{eq:3}-\eqref{eq:actu} yields the same $u(k)$ and $y(k)$ as those of \eqref{eq:cl} for all $k\in\Z_{\geq 0}$
	by Lemma~\ref{lem:perturb}.
\end{proof}
\renewcommand\qedsymbol{$\square$}

Theorem~\ref{thm} implies that given any $\delta>0$, there exist $L$ and $s$ such that $\epsilon\left(L,s\right)<\delta$.
Thus,
given an arbitrary performance error $\epsilon\left(L,s\right)\geq 0$,
Theorem~\ref{thm} provides a guidance to choose proper parameters to construct the encrypted controller \eqref{eq:3}-\eqref{eq:actu} which guarantees the desired performance \eqref{eq:thm}.

\begin{remark}\upshape\label{remark1}
	(\textit{Guide for parameter design})
	The design parameters
	can be selected through the following procedure.
	Once the desired performance error $\epsilon\left(L,s\right)$ is set,
	one is able to choose $L$ and $s$ based on \eqref{eq:thm} and \eqref{eq:epsilon}.
	Next, the plaintext space size $N$ is determined by \eqref{eq:Nbound}.
	Meanwhile, the parameter $\bar{r}$ of Property~\ref{prop:1} should satisfy $\bar{r}\geq \bar{n}$ due to the homomorphic operations in $\eqref{eq:3}$.
	Given $N$ and $\bar{r}$, the remaining parameters of the HE scheme in use are determined
	so that the desired security level is achieved and
	Property~$\ref{prop:1}$ holds.\qed
\end{remark}

\section{Customized Design for RLWE-based Cryptosystems}\label{sec:rlwe}

This section proposes a customized design of the encrypted controller presented in Section~\ref{sec:swhe} for RLWE-based cryptosystems,
utilizing the properties introduced in Section~\ref{subsec:rlwe} where
the addition and multiplication of multiple messages can be computed at once.
Consequently,
the proposed method
reduces
the number of homomorphic operations at each time step,
the communication load between the plant and the controller,
and the amount of encrypted control parameters stored at the controller compared to the design in Section \ref{sec:swhe}.

In order to utilize Property~\ref{prop:2},
we encrypt the plant input $u(k)$ and the output $y(k)$ each into a single ciphertext,
then represent \eqref{eq:uz} as a combination of element-wise additions and multiplications.
To begin with, let the matrix $\mathcal{H}$ be split into $2n$ matrices
as
\begin{align*}
	\mathcal{H}=\left[\begin{array}{ccc|ccc}
		\smash[b]{\underbrace{\strut\mathcal{H}_1}_{l}} & \cdots & \smash[b]{\underbrace{\strut\mathcal{H}_n}_{l}} & \smash[b]{\underbrace{\strut\mathcal{H}_{n+1}}_{h}} & \cdots & \smash[b]{\underbrace{\strut\mathcal{H}_{2n}}_{h}}
	\end{array}\right]\in\R^{h\times n\left(l+h\right)},
\end{align*}\\
where $\mathcal{H}_i\in\R^{h\times l}$ and $\mathcal{H}_{n+i}\in\R^{h\times h}$, $i=1,2,\ldots,n$.
Then, \eqref{eq:zdyn} can be rewritten as
\begin{align}\label{eq:sum}
    u(k)=\sum_{i=1}^n \mathcal{H}_iy(k-i)+\mathcal{H}_{n+i}u(k-i).
\end{align}
Now we need to express each matrix-vector multiplication by element-wise operations between vectors.

\begin{figure}[t!]
	\centering
	\includegraphics[width=\columnwidth]{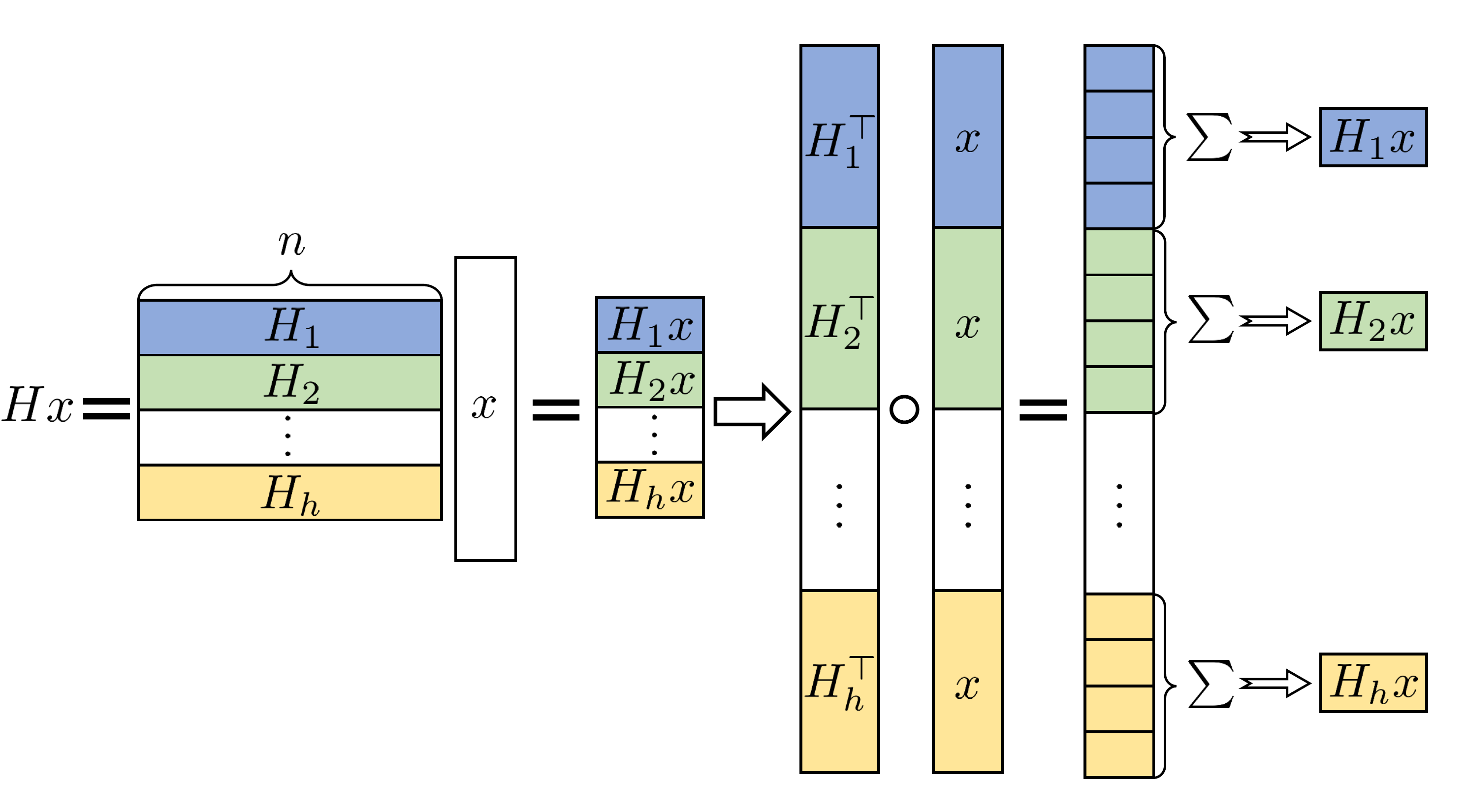}
	\caption{Product $Hx$ implemented with Hadamard product, where each row of $H$ is denoted by $H_i$ for $i=1,\,2,\,\ldots,\,h$.}
	\label{fig:slotmult}
\end{figure}

Consider, for example, the product $Hx$ between a matrix $H\in \R^{h\times n}$ and a vector $x\in \R^n$, as shown in Fig.~\ref{fig:slotmult}, where the $i$-th row of $H$ are denoted by $H_i\in\R^{1\times n}$.
In Fig.~\ref{fig:slotmult}, the matrix $H$ is vectorized to be a column vector of length $hn$, and the vector $x$ is duplicated $h$ times to build another vector of the same length.
Such column vectors of length $hn$ can be regarded as having $h$ ``partitions'' of length $n$.
When these two vectors of length $hn$ are multiplied element-wisely,
we obtain another vector of length $hn$ whose elements in the $i$-th partition are summed up to be $H_ix$, \textit{i.e.}, the $i$-th element of $Hx$.
The overall process can be summarized as
\begin{align}\label{eq:mtxmult}
	Hx=\col\left\lbrace \left\langle\vect( H^\top)\circ \left(\mathbf{1}_h\otimes x\right),\,e_i\otimes\mathbf{1}_n\right\rangle\right\rbrace _{i=1}^h,
\end{align}
where $e_i\in\R^h$ is a unit vector whose only nonzero element is its $i$-th element.

\begin{figure}[t!]
	\centering
	\includegraphics[width=\columnwidth]{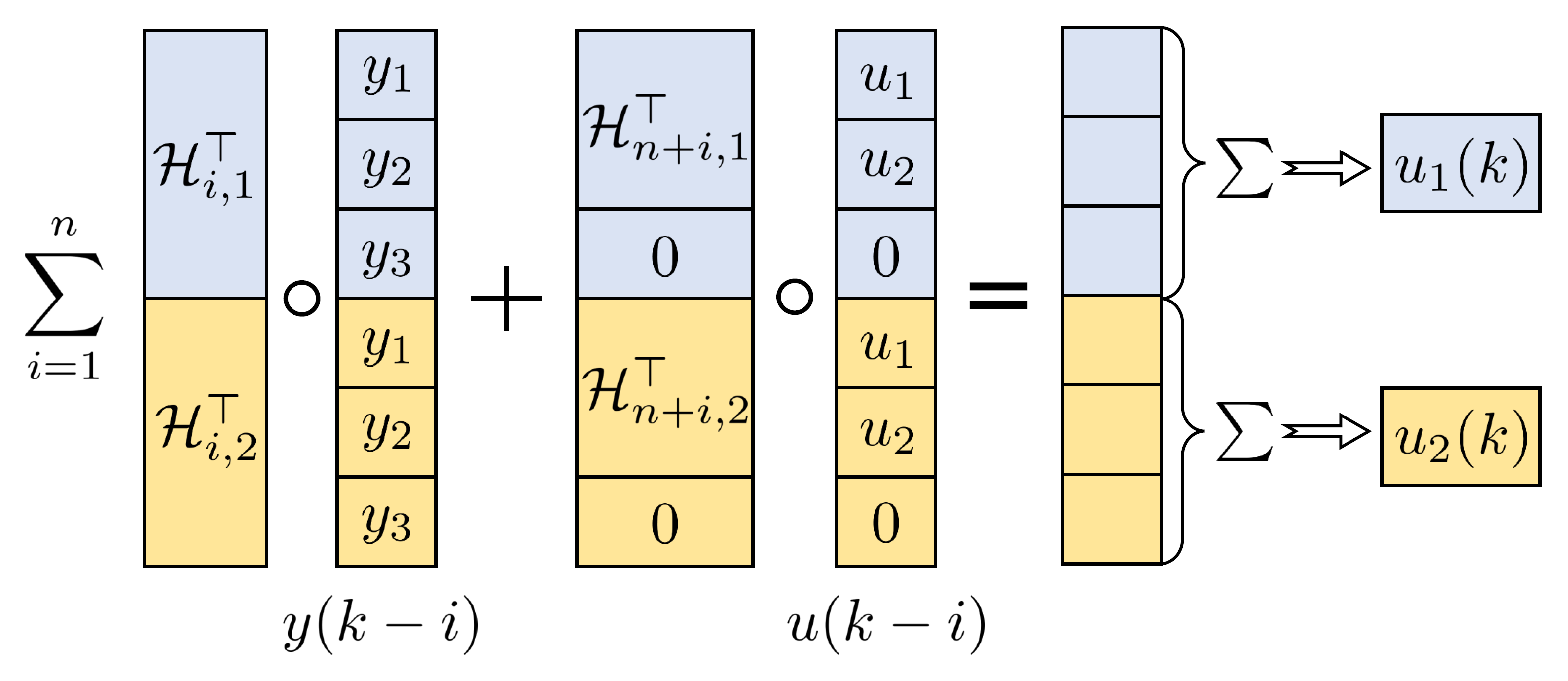}
	\caption{Implementation of (\ref{eq:uz}) using (\ref{eq:mtxmult}) when $l=3$, $h=2$, and $p=6$, where $\mathcal{H}_{i,j}$ is the $j$-th row of $\mathcal{H}_i$ for $i=1,\,2,\,\ldots,\,2n$.}
\label{fig:slotsum}
\end{figure}

The method of \eqref{eq:mtxmult} is then applied to matrix-vector multiplications in \eqref{eq:sum}, as shown in Fig.~\ref{fig:slotsum},
by duplicating the plant output and input $h$-times to build column vectors of length $h\cdot\max\lbrace h,l\rbrace$.
When $h\neq l$, zeros are padded in these vectors of length $h\cdot\max\lbrace h,l\rbrace$ to fit the length of each partition, as in the case depicted in Fig.~\ref{fig:slotsum}.
Then, the summations of elements within each partition are performed only at the end to yield $u(k)$, after element-wise multiplications between $2n$-pairs of vectors followed by element-wise additions.

Note that the summation of elements within each partition cannot be implemented solely by element-wise operations.
Indeed, methods to implement the matrix-vector multiplication over RLWE-based ciphertexts have been developed using an algorithm called ``rotation'' \cite{mtxmult, mtxmult2},
which allows a permutation of elements within a vector.
However, rotation requires additional storage for the encrypted controller,
since a given ciphertext is decomposed and then multiplied by the ``rotation key'' which should be stored in the controller\footnote{For numerical analysis on the effect of applying rotation, see Section~\ref{subsec:previous}.} \cite{bgv, bfv}.

Instead, we make use of the fact that the output of the encrypted controller is re-encrypted at the actuator, as in \eqref{eq:actu},
so that the elements can be summed up at the actuator after decryption.
This is reasonable considering that the actuator is capable of decryption, which already involves addition of scalars.
Therefore, we encrypt the controller \eqref{eq:sum} to operate
\begin{multline}\label{eq:op}
    \sum_{i=1}^n\left( \vect(\mathcal{H}_i^\top)\circ \left(\mathbf{1}_h\otimes y(k-i)\right)\right.\\ \left.+\vect(\mathcal{H}_{n+i}^\top)\circ \left(\mathbf{1}_h\otimes u(k-i)\right)\right),
\end{multline}
and leave the rest of the operation to the actuator.

In \eqref{eq:op}, it is assumed that $h=l$, which is assumed for the rest of this section to ignore the padded zeros for simplicity.
In addition,
we assume that $p=h^2$, and hence the vectors in Fig.~\ref{fig:slotsum} are fully packed into each plaintext in $R_{p,N}$.
Otherwise, zeros are again padded at the end of each vector before packing.

Now each 
$\mathcal{H}_i$ is vectorized as the matrix $H$ in \eqref{eq:mtxmult}, then quantized and encrypted
analogously to \eqref{eq:boldy};
\begin{align}\label{eq:Hi}
	\mathbf{H}_i:=\Enc^\prime\left( \Ceil \frac{\vect( \mathcal{H}_i^\top)} {s}\Flr\right),\quad
	\text{for}
	\,\,\,i=1,\,2,\,\ldots,\,2n,
\end{align}
where $1/s\geq 1$ is again a scaling parameter and
$\Enc^\prime$, the composition of encryption and packing defined in Section~\ref{subsec:rlwe}, is used.
In order to be multiplied with each $\mathbf{H}_i$,
the plant output and input are duplicated $h$ times, quantized, and encrypted as
\begin{equation}\label{eq:sens}
    \begin{aligned}
		\mathbf{y}(k)&=\Enc^\prime\left(\Ceil \frac{\mathbf{1}_h\otimes y(k)}{L}\Flr\right),\\
		\mathbf{u}(k)&=\Enc^\prime\left(\Ceil\frac{\mathbf{1}_h\otimes u(k)}{L} \Flr\right),
    \end{aligned}
\end{equation}
respectively at the sensor and the actuator
with $1/L> 0$.

The encrypted controller operates \eqref{eq:op} over the encrypted parameters \eqref{eq:Hi} and the inputs \eqref{eq:sens} transmitted from the sensor and the actuator, then returns the output $\bar{\mathbf{u}}(k)$
following the procedure described in Algorithm~\ref{alg}.
The initial condition of the encrypted controller is also set by $z_0$ in Lemma~\ref{lem:M},
but it is transformed into the duplicated form as in \eqref{eq:z0}.
Regarding the dynamics \eqref{eq:zdyn} of $z(k)$, the operation of \eqref{eq:uz} is implemented as \eqref{eq:4},
utilizing the homomorphic operation $\Prod_2$ in Property~\ref{prop:2}, and
the state update \eqref{eq:z} is implemented as
Steps~\ref{state:z1} and \ref{state:z2} of Algorithm~\ref{alg}.

\begin{algorithm}[t!]
	\caption{Encrypted controller design customized for RLWE-based cryptosystems.}\label{alg}
	\begin{algorithmic}[1]
		\renewcommand{\algorithmicrequire}{\textbf{Setup:}}
		\REQUIRE Let $z_0=:\col\left\lbrace z_0^i\right\rbrace_{i=1}^{2n}$, where each $z_0^i\in\R^h$. Define
		\begin{align}\label{eq:z0}
			\mathbf{z}_i(k):=\Enc^\prime\left( \Ceil\frac{\mathbf{1}_h\otimes z_0^i}{L}\Flr\right) 
		\end{align}
		for $i=1,\,2,\,\ldots,\,2n$ and set $k=0$.
		\renewcommand{\algorithmicrequire}{\textbf{Input:}}
		\REQUIRE $\mathbf{u}(k)$ and $\mathbf{y}(k)$
		\STATE Compute $\bar{\mathbf{u}}(k)$ as
		\begin{align}\label{eq:4}
			\bar{\mathbf{u}}(k)=\Prod_2\left( \left\lbrace \mathbf{H}_i\right\rbrace_{i=1}^{2n}, \left\lbrace \mathbf{z}_i(k)\right\rbrace_{i=1}^{2n} \right).
		\end{align}
		\STATE Set $\mathbf{z}_1(k+1)= \mathbf{y}(k)$ and $\mathbf{z}_{n+1}(k+1)=\mathbf{u}(k)$.\label{state:z1}
		\STATE For $i=2,\,3,\,\ldots,\,n$, set\label{state:z2}
		\begin{align*}
			\mathbf{z}_i(k+1)= \mathbf{z}_{i-1}(k)\quad \text{and}\quad
			\mathbf{z}_{n+i}(k+1)= \mathbf{z}_{n+i-1}(k).
		\end{align*}
		\STATE Update $k\gets k+1$.
		\renewcommand{\algorithmicensure}{\textbf{Output:}}
		\ENSURE $\bar{\mathbf{u}}(k)$
	\end{algorithmic}
\end{algorithm}

After
the encrypted controller output $\bar{\mathbf{u}}(k)$ is
transmitted to the actuator,
it is decrypted, re-scaled, and processed as
\begin{align}\label{eq:act}
	u(k)=\col\left\lbrace \left\langle \Dec^\prime\left(\bar{\mathbf{u}}(k)\right)\cdot Ls,\,e_i\otimes \mathbf{1}_h \right\rangle\right\rbrace_{i=1}^h,
\end{align}
where the elements within the $i$-th partition of $\Dec^\prime(\bar{\mathbf{u}}(k))\cdot Ls$ are summed up to be the $i$-th element of $u(k)$, as in \eqref{eq:mtxmult}.

The following theorem is analogous to Theorem~\ref{thm} so that the performance error,
in terms of the difference between the control input $u(k)$ generated by the encrypted controller and $u^\prime(k)$ of the original controller \eqref{eq:ctr},
is assured to be under a certain bound,
which can be made arbitrarily small by adjusting the quantization parameters $L$ and $s$.

\begin{theorem}\upshape\label{thm2}
	There exists\footnote{See (\ref{eq:epsilon}) in the proof of Theorem \ref{thm} for an explicit form of $\left\lbrace \epsilon_0,\,\epsilon_1,\,\epsilon_2,\,\epsilon_3\right\rbrace $.} a set of positive numbers
	$\left\lbrace \epsilon_0,\,\epsilon_1,\,\epsilon_2,\,\epsilon_3\right\rbrace $
	such that the encrypted controller of \eqref{eq:sens}, \eqref{eq:act}, and Algorithm~\ref{alg} guarantees \eqref{eq:thm} for all $k\in\Z_{\geq 0}$,
	provided that $N$, $L$, and $s$ satisfy
	\begin{multline}\label{eq:N2}
			\left( \frac{1}{L}
				\max\left\lbrace \epsilon(L,s)+S,\,\lVert z_0\rVert\right\rbrace
			+\frac{1}{2} \right)\\
			\cdot\left(\frac{1}{s}\left\lVert \vect\left( \sum_{i=1}^{2n}\mathcal{H}_i^\top\right) \right\rVert+n \right)
			<\frac{N}{2}
	\end{multline}
	and $1/s>\epsilon_0$.\qed
\end{theorem}

\renewcommand\qedsymbol{$\blacksquare$}
\begin{proof}
First, we follow the proof of Lemma~\ref{lem:perturb} analogously
	and consider the controller \eqref{eq:ze}.
	Suppose that $u(k)$ and $y(k)$ of \eqref{eq:ze} are bounded as
	\begin{multline}\label{eq:pfN}
		\left(\frac{1}{L}\max\left\lbrace \left\lVert u(k)\right\rVert,\left\lVert y(k)\right\rVert, \lVert z_0\rVert \right\rbrace+\frac{1}{2}  \right)\\
		\cdot\left( \frac{1}{s}\left\lVert \vect\left(\sum_{i=1}^{2n}\mathcal{H}_i^\top \right) \right\rVert+n\right)<\frac{N}{2} 
	\end{multline}
	for $k=0,\,1,\,\ldots,\,T$ with some $T\in\N$.
	By induction, it can be shown that for $k=0,\,1,\,\ldots,\,T$,
	\begin{align*}
		z(k)=L\cdot\col\left\lbrace
		\begin{bmatrix}
			I_h & \mathbf{0}_{h\times \left(p-h\right)}
		\end{bmatrix}
		\Dec^\prime\left( \mathbf{z}_i(k)\right) \right\rbrace_{i=1}^{2n},
	\end{align*}
	since for all $k\in\Z_{\geq 0}$,
	\begin{align*}
			\mathbf{z}_i(k)=\begin{cases}
				\Enc^\prime\left(\Ceil  \left(\mathbf{1}_h\otimes y(k-i)\right)/L\Flr \right) & \text{for}\,\,i\in\left[ 1,n\right] ,\\
				\Enc^\prime \left( \Ceil \left(\mathbf{1}_h\otimes u(k-i)\right)/L \Flr\right) & \text{for}\,\,i\in\left[ n+1,2n\right]\\
			\end{cases}
	\end{align*}
	by Algorithm~\ref{alg}, and
	\begin{align*}
		\Dec^\prime\left( \bar{\mathbf{u}}(k)\right)
		= \sum_{i=1}^{2n}\vect\left( \Ceil \frac{\mathcal{H}_i}{s}\Flr^\top\right)\circ\Dec^\prime\left(\mathbf{z}_i(k) \right) \,\,\,\mathrm{mod}\,\,N
	\end{align*}
	by Property~\ref{prop:2}.
	Note that \eqref{eq:pfN} ensures
	\begin{align*}
		\left\lVert \sum_{i=1}^{2n}\vect\left( \Ceil \frac{\mathcal{H}_i}{s}\Flr^\top\right)\right\rVert\cdot\left\lVert z(k)\right\rVert<\frac{N}{2}.
	\end{align*}
	Thus, the encrypted controller of \eqref{eq:sens}, \eqref{eq:act}, and Algorithm~\ref{alg} generates the same control input $\left\lbrace u(k) \right\rbrace_{k=0}^T $ as \eqref{eq:ze},
	which can be transformed to the perturbed controller \eqref{eq:xe} satisfying \eqref{eq:eubound}.
	The rest of the proof is analogous to the proof of Theorem~\ref{thm}.
	The only difference is that the parameters satisfy \eqref{eq:N2} which is derived directly from \eqref{eq:pfN}.
\end{proof}
\renewcommand\qedsymbol{$\square$}

Note that the condition \eqref{eq:N2} of Theorem~\ref{thm2} differs from \eqref{eq:Nbound} of Theorem~\ref{thm}.
This is because unlike \eqref{eq:actu} where the encrypted controller output contains the message of $u(k)$,
the output of Algorithm~\ref{alg} has the outcome of element-wise operations in \eqref{eq:4} as its message.
Thus, the condition \eqref{eq:N2} ensures that every element of each partition inside $\bar{\mathbf{u}}(k)$ belongs to $\Z_N$.

The design parameters
can be determined through a process analogous to 
Remark~\ref{remark1}.
The difference is that
after the parameters $L$ and $s$ are determined, the size of the plaintext space $N$ is chosen to satisfy \eqref{eq:N2}, instead of \eqref{eq:N}.
The parameter $\bar{r}$ of Property~\ref{prop:2} should also be greater than or equal to $2n$, the number of homomorphic additions in \eqref{eq:4}.
Moreover, in order for
the duplicated vectors in \eqref{eq:sens} to be packed inside
the plaintexts of RLWE-based cryptosystems,
the parameter $p$ needs to be at least
$h\cdot\max\left\lbrace h,\,l\right\rbrace$.
Given these requirements,
one can refer to \cite{HEstandard} in choosing an appropriate pair of $p$ and $q$ that achieves the desired level of security.

\section{Discussions}\label{sec:discussion}

\begin{table}[t]
	\caption{Comparison of Encrypted Controllers in Sections~\ref{sec:swhe} and \ref{sec:rlwe} Implemented by RLWE-based Cryptosystems}
	\label{table}
	\begin{center}
 \renewcommand{\arraystretch}{1.2}
		\begin{tabular}{c|c||c|c}
			\hline  \multicolumn{2}{c||}{}   & Section~\ref{sec:swhe} & Section~\ref{sec:rlwe}\\
			\hline
			& $\Enc$ & $h+l$ & $2$\\
			\# of operations & $\Dec$ & $h$ & $1$\\
			executed at each $k$ & $\oplus$ & $h\left(nh+nl-1\right)$ & $2n-1$\\
			& $\Mult$ & $hn\left(h+l\right)$ & $2n$\\ \hline
			& $\mathbf{u}(k)$ & $2h$ & $2$\\
			& $\bar{\mathbf{u}}(k)$ & $3h$ & $3$\\
			\# of polynomials in& $\mathbf{y}(k)$ & $2l$ & $2$\\
			& $\mathbf{z}(k)$ & $2n\left(h+l\right)$ & $4n$ \\
			& $\mathbf{H}$ & $2hn\left(h+l\right)$ & $4n$ \\
			\hline
		\end{tabular}
	\end{center}
\end{table}

\begin{table}[!t]
\caption{Comparison of Encrypted Controller in Section~\ref{sec:rlwe} with Previous Results}
	\label{tableB}
	\begin{center}
\renewcommand{\arraystretch}{1.2}
		\begin{tabular}{c||c|c}
		\hline	At each $k$, & computation load & communication load \\ \hline
            Algorithm~\ref{alg} & $O(np\log p)$ & $7p$\\
            \cite{Teranishi} & $O((n+h+l)dp\log p)$ & $6p$\\ 
            \cite{Kim22TAC} & $O(n(n+h+l)dp^2)$ & $(2h+l)(p+1)$ \\ \hline
		\end{tabular}
	\end{center}
\end{table}

\subsection{Effect of Customization in Section~\ref{sec:rlwe}}

We have proposed two approaches to encrypt linear dynamic controllers; the first approach, discussed in Section~\ref{sec:swhe}, is designed to be adaptable to a wide range of HE schemes, whereas the second approach, presented in Section~\ref{sec:rlwe}, strategically leverages the features of RLWE-based cryptosystems.
This subsection compares these two approaches in terms of the computation load and the storage occupied by each encrypted data, as summarized in Table~\ref{table}.
For comparison, it is assumed that both controllers are implemented using RLWE-based cryptosystems.
To examine the computation load,
we have calculated the number of encryptions, decryptions, and homomorphic operations performed at each time step.
We have also counted the number of polynomials composing each encrypted data, in order to analyze the storage consumption\footnote{Recall that each component of $\bar{\mathbf{u}}(k)$ belongs to $R_{p,q}^3$, whereas other ciphertexts belong to $R_{p,q}^2$.}.
The communication load is determined by the storage consumption of $\mathbf{u}(k)$, $\bar{\mathbf{u}}(k)$, and $\mathbf{y}(k)$.

In conclusion,
the customization proposed in Section~\ref{sec:rlwe} is more efficient than
the general design presented in Section~\ref{sec:swhe},
especially when the plant is a multi-input multi-output system where either $h$ or $l$ is larger than $1$.
Specifically, the efficiency of the customized design comes from the fact that the number of homomorphic multiplications executed at each time step depends only on the order $n$ of the controller, being independent from $h$ and $l$.

\subsection{Comparison to Previous Results}\label{subsec:previous}

The proposed controller in Section~\ref{sec:rlwe} is compared with two previous results;
in \cite{Teranishi}, a RLWE-based cryptosystem is utilized but the matrix-vector multiplication is implemented differently,
and in \cite{Kim22TAC}, an LWE-based cryptosystem is utilized with the external product \cite{gsw} between ciphertexts.
We have analyzed the computation and communication load of these three encrypted controllers, as shown in Table~\ref{tableB}.
For a fair comparison, the parameters of the LWE-based cryptosystem are set as follows: the plaintext space is $\Z_N$, the length of the secret key\footnote{It is a key essential for encryption and decryption in a cryptosystem.} is $p$, and each ciphertext consists of integers in $\Z_q$.

The computation load is examined in terms of the number of scalar multiplications executed at each time step by these encrypted controllers.
As shown in Table~\ref{table}, our proposed controller in Section~\ref{sec:rlwe} performs $2n$-homomorphic multiplications.
Each of these homomorphic multiplications requires $O(p\log p)$-scalar multiplications,
since it accompanies $4$-polynomial multiplications, as described in \cite{bfv} and \cite{bgv},
and a polynomial multiplication can be computed through $O(p\log p)$-scalar multiplications \cite{NTT}.
As a result, the proposed controller conducts $O(np\log p)$-scalar multiplications.

The controller of \cite{Teranishi} executes $(n+1)$-homomorphic multiplications,
followed by the same number of relinearizations,
and $(h+l-1)$-rotations.
Relinearization is for reducing the number of polynomials in each ciphertext from $3$ to $2$ after homomorphic multiplications,
and rotation is utilized to implement matrix-vector multiplications over RLWE ciphertexts.
Meanwhile, the controller proposed in \cite{Kim22TAC} executes $(n^2+2hn+ln)$-external products at each time step,
since the matrices and signals are encrypted element-wisely.

Relinearization, rotation, and external product have in common that they decompose a given ciphertext in base $\nu\in \N$, where $\nu$ is usually far smaller than $q$, and compute over this expanded ciphertext.
Both relinearization and rotation carry $2d$-polynomial multiplications, where $d:=\lfloor \log_\nu q \rfloor$ \cite{bfv, bgv}.
The external product of LWE-based cryptosystems multiplies a vector of length $d(p+1)$ with a $(p+1)\times d(p+1)$ matrix \cite{gsw}, which corresponds to $d(p+1)^2$-scalar multiplications.
Hence, the total number of scalar multiplications performed by each encrypted controller can be derived as in Table~\ref{tableB}.

The communication load is
computed as the number of integers transmitted between the plant and the controller at each time step.
Our proposed controller in Section~\ref{sec:rlwe} receives $\mathbf{y}(k)\in R_{p,q}^2$ and $\mathbf{u}(k)\in R_{p,q}^2$, then returns $\mathbf{\bar{u}}(k)\in R_{p,q}^3$.
Thus, the overall communication load is $2p+2p+3p=7p$.
On the other hand, the controller of \cite{Teranishi} returns a ciphertext in $R_{p,q}^2$ thanks to relinearization, and hence the communication load is $2p+2p+2p=6p$.
Unlike the other two controllers, every signal is encrypted and computed element-wisely in \cite{Kim22TAC}.
That is, the controller receives $(h+l)$-ciphertexts and transmits $h$ ciphertexts to the plant.
Since a ciphertext of LWE-based cryptosystems belongs to $\Z_q^{p+1}$ \cite{lwe},
it carries $(2h+l)(p+1)$-amount of communication load.

It can be observed from Table~\ref{tableB} that the proposed controller
in Section~\ref{sec:rlwe}
requires less amount of computations compared to the methods of \cite{Teranishi} and \cite{Kim22TAC}.
Although the communication load of our design is greater than that of \cite{Teranishi},
it can also be reduced to $6p$ by applying relinearization, which increases the number of scalar multiplications to $O((n+d)p\log p)$.
Still, the computation load is less than that of \cite{Teranishi}.

\section{Simulation Results}\label{sec:simul}
This section provides simulation results of the proposed method in Section \ref{sec:rlwe} applied to a controller
stabilizing the model of AFTI/F-16 aircraft \cite{simul2};
by discretizing the continuous-time plant, we have the plant \eqref{eq:plant} with matrices
\begin{align}\label{eq:simulplant}
    A&={\small\begin{bmatrix}
		1.0000 &   0.0020  &  0.0663 &   0.0047   & 0.0076\\
		0  &  1.0077  &  2.0328  & -0.5496  & -0.0591\\
		0  &  0.0478  &  0.9850 &  -0.0205 &  -0.0092\\
		0    &     0     &    0  &  0.3679     &    0\\
		0    &     0    &     0     &    0  &  0.3679
	\end{bmatrix}},\notag\\
    B&={\small\begin{bmatrix}
		0.0029  &  0.0045\\
		-0.3178 &  -0.0323\\
		-0.0086  & -0.0051\\
		0.6321    &     0\\
		0  &  0.6321
	\end{bmatrix}},\\
    C&={\small\begin{bmatrix}
		0 &   1     &    0      &   0    &     0\\
		0  & -0.2680  & 47.7600  & -4.5600  &  4.4500\\
		1     &    0     &    0     &    0    &     0\\
		0     &    0     &    0  &  1     &    0\\
		0    &     0     &    0    &     0  &  1
	\end{bmatrix}},\notag
\end{align}
under the sampling period $0.05$s.
The initial state of the plant is set as $x_p(0)=\left[1,\,-1,\,0,\,0.7,\,1 \right]^\top $.
Let the controller \eqref{eq:ctr} be designed as
\begin{align*}
	x(k+1)&=\left(A-L_cC+BK_c\right)x(k)\!+\!L_cy(k),\\
	u(k)&=K_cx(k),
\end{align*}
which is in the observer-based form with the gains
\begin{align*}
	L_c&:={\small \begin{bmatrix}
		0.0011 &   0.0014 &   0.5868  &  0.0056  &  0.0007\\
		0.6296 &   0.0429 &  -0.0003 &  -0.1811 &  -0.1278\\
		0.0326 &   0.0205  &  0.0000  &  0.0337  & -0.0480\\
		-0.0049  & -0.0003  &  0.0002 &   0.1732 &   0.0005\\
		-0.0037 &   0.0003  &  0.0000  &  0.0005  &  0.1733
	\end{bmatrix}},\\
	K_c&:={\small \begin{bmatrix}
		0.5743 &   0.5544  &  3.6332 &  -0.3636 &  -0.0668\\
		-1.8788 &  -0.3166 &  -2.3100  &  0.2151  &  0.0691
	\end{bmatrix}},
\end{align*}
and the initial state $x(0)=\left[ -0.001,\,0.013,\,0.2,\,-0.02,\,0\right]^\top $.

We implemented the encrypted controller of \eqref{eq:sens}, \eqref{eq:act}, and Algorithm~\ref{alg} using Lattigo \cite{lattigo}, an HE library which supports RLWE-based cryptosystems in Go.
The proposed controller was encrypted by the BGV scheme \cite{bgv}
with the encryption parameters set as follows;
$N=65929217\approx 2^{26}$ and $p=2^{12}$ for the plaintext space $\mathcal{R}_{p,N}$, $q=18889455798646780911617\approx 2^{74}$ for the ciphertext space $\mathcal{R}_{p,q}$, and the standard deviation of the error distribution\footnote{See Appendix~B for the precise meaning of this distribution.} $\sigma = 3.2$.
The parameters were selected to ensure $128$-bit security \cite{HEstandard} and satisfy Property~\ref{prop:2} for $\bar{r}\geq 2n$.

\begin{figure}[t!]
	\centering
%
%
\begin{tikzpicture}

\begin{axis}[%
width=0.4\textwidth,
height=0.15\textheight,
at={(0.88in,0.582in)},
scale only axis,
xmin=0,
xmax=5,
xlabel style={font=\color{black}},
xlabel={Time (s)},
ymin=0,
ymax=0.05,
ylabel style={font=\color{black}},
ylabel={$|| u(k)-u^\prime(k)||$},
axis background/.style={fill=white},
xmajorgrids,
ymajorgrids,
legend style={legend cell align=left, align=left, draw=black},
ytick = {0.01,0.02,0.03,0.04}
]
\addplot [color=black, line width=1.6pt]
  table[row sep=crcr]{%
0	0.000174585022267109\\
0.05	0.000145624742475426\\
0.1	0.000252776667355952\\
0.15	0.000815780842672177\\
0.2	0.000746451751419719\\
0.25	0.000460419782954182\\
0.3	0.00101055296459451\\
0.35	0.00115082529618066\\
0.4	0.00150277761700538\\
0.45	0.00214369291233251\\
0.5	0.00234453936945138\\
0.55	0.00303122674185291\\
0.6	0.00341719639777915\\
0.65	0.0034312904788466\\
0.7	0.00391825603285086\\
0.75	0.00402637539432761\\
0.8	0.00398472036262889\\
0.85	0.00397510861896765\\
0.9	0.00416466037849621\\
0.95	0.00459594317392018\\
1	0.00478176256828573\\
1.05	0.0044235835768011\\
1.1	0.00451865755381199\\
1.15	0.00470509678928649\\
1.2	0.00480565872326995\\
1.25	0.00459455521958829\\
1.3	0.00413562991793819\\
1.35	0.00392555179152309\\
1.4	0.00425267962386468\\
1.45	0.00378212453772917\\
1.5	0.00392645601157614\\
1.55	0.00375900044939621\\
1.6	0.00346995732581582\\
1.65	0.00345973997796959\\
1.7	0.00362745201888975\\
1.75	0.00361061952065014\\
1.8	0.00333281883383086\\
1.85	0.00289707107553449\\
1.9	0.00266103602240141\\
1.95	0.00266615162980544\\
2	0.00285246193608051\\
2.05	0.00309516337020978\\
2.1	0.00293208198648209\\
2.15	0.0027495014813434\\
2.2	0.0027462097278524\\
2.25	0.00251724525158062\\
2.3	0.00242503896521439\\
2.35	0.00201949673259649\\
2.4	0.00231115875634938\\
2.45	0.00202868863502565\\
2.5	0.00191130783479522\\
2.55	0.00209349710801064\\
2.6	0.00213221736152576\\
2.65	0.00190203017928139\\
2.7	0.00185284457116356\\
2.75	0.00179396425446739\\
2.8	0.0019963284478839\\
2.85	0.00144864340857312\\
2.9	0.0016591511884293\\
2.95	0.0015105802274851\\
3	0.0016972311189177\\
3.05	0.00174547622726136\\
3.1	0.00160490596423043\\
3.15	0.00139564206471283\\
3.2	0.00129099212639014\\
3.25	0.00129089349226631\\
3.3	0.00111467011924482\\
3.35	0.000960754811752763\\
3.4	0.00110827305959599\\
3.45	0.00127616414690395\\
3.5	0.000936352251268881\\
3.55	0.000684108493120849\\
3.6	0.00109157747185687\\
3.65	0.000997825584699985\\
3.7	0.000759317152463689\\
3.75	0.000862837580569189\\
3.8	0.00112591394036377\\
3.85	0.000747158198769555\\
3.9	0.000612566004974195\\
3.95	0.000543502519532505\\
4	0.00044954841760669\\
4.05	0.000724089721714154\\
4.1	0.000740851580637462\\
4.15	0.000559997040133407\\
4.2	0.000809215755144535\\
4.25	0.000593612420438779\\
4.3	0.000765791512084049\\
4.35	0.000994121139853516\\
4.4	0.000727854198341654\\
4.45	0.000597768607334872\\
4.5	0.000734330243198205\\
4.55	0.000660775385397993\\
4.6	0.000750991157381123\\
4.65	0.000668530990142755\\
4.7	0.000748424771727844\\
4.75	0.000522911880985085\\
4.8	0.000419911841724201\\
4.85	0.000714964501505895\\
4.9	0.000631823854083197\\
4.95	0.000515260949525655\\
};
\addlegendentry{\small $1/L=2\times 10^3,\,1/s=10^4$}

\addplot [color=blue, dashed, line width=1.6pt]
  table[row sep=crcr]{%
0	0.00274250225341753\\
0.05	0.000898626046902119\\
0.1	0.00178689230131848\\
0.15	0.000851959353046952\\
0.2	0.00260970660698234\\
0.25	0.00360576190121094\\
0.3	0.00204636261416908\\
0.35	0.0016363145169833\\
0.4	0.00259834372108191\\
0.45	0.00219207491159945\\
0.5	0.0027051450633724\\
0.55	0.00390079966151594\\
0.6	0.00247195470758121\\
0.65	0.00374732786363044\\
0.7	0.00493848328227566\\
0.75	0.00873974894374426\\
0.8	0.00416448748262876\\
0.85	0.00612799744343388\\
0.9	0.00219203208550982\\
0.95	0.00234349494300727\\
1	0.000964938120734669\\
1.05	0.00420777806385394\\
1.1	0.00675580300968834\\
1.15	0.00783556890707339\\
1.2	0.00474320543507235\\
1.25	0.00195850953550446\\
1.3	0.00271392141046462\\
1.35	0.00374372830319497\\
1.4	0.00519385081331305\\
1.45	0.00445216726194864\\
1.5	0.00421409518646743\\
1.55	0.00519305347613783\\
1.6	0.002337363911481\\
1.65	0.00292795933769936\\
1.7	0.00417067910970075\\
1.75	0.00400671904593792\\
1.8	0.0022191112154309\\
1.85	0.00128449328303955\\
1.9	0.00099164679587939\\
1.95	0.00181999561493625\\
2	0.000326931833476047\\
2.05	0.00252220509379691\\
2.1	0.0014505132533764\\
2.15	0.000166150913012874\\
2.2	0.00211164648049159\\
2.25	0.00248782113444321\\
2.3	0.00380839502685232\\
2.35	0.00235055845756787\\
2.4	0.00239477918979873\\
2.45	0.00109009712209211\\
2.5	0.00296991425784609\\
2.55	0.00179483480649124\\
2.6	0.00159234301106431\\
2.65	0.00124621120973896\\
2.7	0.00187815313379806\\
2.75	0.00153089969359295\\
2.8	0.00142845425056233\\
2.85	0.0018094597492111\\
2.9	0.000897954080253595\\
2.95	0.00309100502010025\\
3	0.00298115167273048\\
3.05	0.0031291431529737\\
3.1	0.000942928491982723\\
3.15	0.00418341316486175\\
3.2	0.00102157604680954\\
3.25	0.00164702861400349\\
3.3	0.0010484540365584\\
3.35	0.00278211706959797\\
3.4	0.00546563806637699\\
3.45	0.00275808035148035\\
3.5	0.00229480633968391\\
3.55	0.00360584442422014\\
3.6	0.000459603553928443\\
3.65	0.000591354180450844\\
3.7	0.000985130164626378\\
3.75	0.00289629771709339\\
3.8	0.00135852686594045\\
3.85	0.00182610850626502\\
3.9	0.000761983852277961\\
3.95	0.00258249897734466\\
4	0.0036012246241207\\
4.05	0.00162127524079114\\
4.1	0.00258781152104909\\
4.15	0.00262489842211841\\
4.2	0.00104954333503771\\
4.25	0.00111112980189565\\
4.3	0.0019447301985683\\
4.35	0.00368344577132883\\
4.4	0.004166629184579\\
4.45	0.00256158227234317\\
4.5	0.0015208273421676\\
4.55	0.00101848313583599\\
4.6	0.000188819834466643\\
4.65	0.000886794479087328\\
4.7	0.00151869837744444\\
4.75	0.00215557836253911\\
4.8	0.00321177911148238\\
4.85	0.00531147107486814\\
4.9	0.0058916790949826\\
4.95	0.00542667753223522\\
};
\addlegendentry{\small $1/L=2\times 10^2,\,1/s=10^4$}

\addplot [color=red, dashdotted, line width=1.6pt]
  table[row sep=crcr]{%
0	0.00151533560969165\\
0.05	0.000327602558553692\\
0.1	0.00133148081284654\\
0.15	0.00452108551838837\\
0.2	0.0089455788688202\\
0.25	0.0106006653151759\\
0.3	0.0159106918854094\\
0.35	0.0198166884316276\\
0.4	0.0240505180429949\\
0.45	0.0282830482419068\\
0.5	0.0324282890298451\\
0.55	0.0354373973846925\\
0.6	0.0379362202932938\\
0.65	0.039567616805776\\
0.7	0.0405937673910653\\
0.75	0.0411507575279714\\
0.8	0.0415545785707571\\
0.85	0.0412442366406884\\
0.9	0.0410291243036296\\
0.95	0.0403077906358721\\
1	0.0395644203688628\\
1.05	0.0387650342845555\\
1.1	0.037585455700712\\
1.15	0.0364937740191449\\
1.2	0.0356471535083334\\
1.25	0.0348155944193151\\
1.3	0.03350978714078\\
1.35	0.0319739852341165\\
1.4	0.0312741625946348\\
1.45	0.0302624711531053\\
1.5	0.0290439050846285\\
1.55	0.0276629654346287\\
1.6	0.0267419473202439\\
1.65	0.0258942268242319\\
1.7	0.0244829799645073\\
1.75	0.0239609380775829\\
1.8	0.0228206117466925\\
1.85	0.0216769792846945\\
1.9	0.0207822307810066\\
1.95	0.0197064429908198\\
2	0.0191736767687654\\
2.05	0.0182710240320634\\
2.1	0.0175642847139098\\
2.15	0.0169928797863779\\
2.2	0.0162786959893105\\
2.25	0.0158830532525254\\
2.3	0.0154003959688071\\
2.35	0.0145085089113118\\
2.4	0.0139847338400627\\
2.45	0.0132451160946696\\
2.5	0.012732966389523\\
2.55	0.012151126427789\\
2.6	0.0116858074728559\\
2.65	0.0110205221695811\\
2.7	0.0105841562341132\\
2.75	0.0102991315345021\\
2.8	0.0098202793131679\\
2.85	0.00938742556071543\\
2.9	0.00920374904239372\\
2.95	0.00872336309907137\\
3	0.00826294413464517\\
3.05	0.00810610044070839\\
3.1	0.00818205581716724\\
3.15	0.00765274182790723\\
3.2	0.00739652864506382\\
3.25	0.00714543401213282\\
3.3	0.00704230720794494\\
3.35	0.0066434837220829\\
3.4	0.0060906133814309\\
3.45	0.00566938030394836\\
3.5	0.00530131387586381\\
3.55	0.00523312559818267\\
3.6	0.0051994754701782\\
3.65	0.00489532367918909\\
3.7	0.00445884541505311\\
3.75	0.00461149285012775\\
3.8	0.00454821083949074\\
3.85	0.00437807341240429\\
3.9	0.00427181948075635\\
3.95	0.00429124541884222\\
4	0.00383627052318197\\
4.05	0.00371312473567973\\
4.1	0.00329440383179771\\
4.15	0.00347625590220289\\
4.2	0.00323130561250479\\
4.25	0.00280908546216044\\
4.3	0.00293541658715376\\
4.35	0.00296542585054081\\
4.4	0.00277020565770876\\
4.45	0.00253527978900697\\
4.5	0.00247974355834865\\
4.55	0.00237361357686191\\
4.6	0.00240089376133896\\
4.65	0.00245719935628645\\
4.7	0.00247215206811215\\
4.75	0.00207892992322001\\
4.8	0.00190242355246184\\
4.85	0.00196031361959266\\
4.9	0.00166624308094854\\
4.95	0.0017271379520997\\
};
\addlegendentry{\small $1/L=2\times 10^3,\,1/s=10^3$}

\end{axis}
\end{tikzpicture}%
	\caption{Performance error $\lVert u(k)-u^\prime(k) \rVert$ of the proposed encrypted controller customized for RLWE-based cryptosystem.}
	\label{fig:simul}
\end{figure}
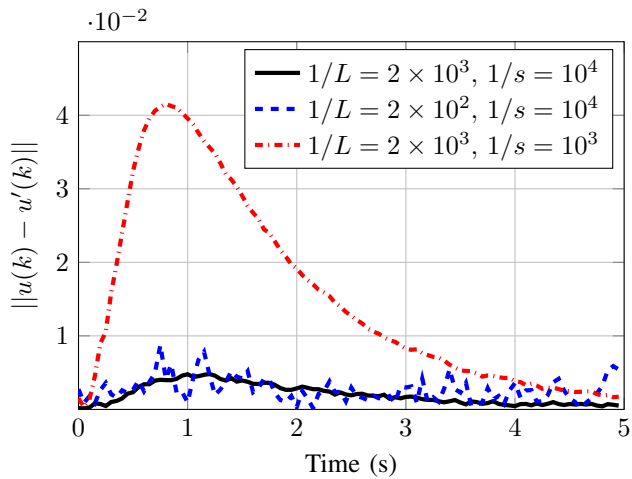

\begin{figure}[t!]
	\centering
	\input{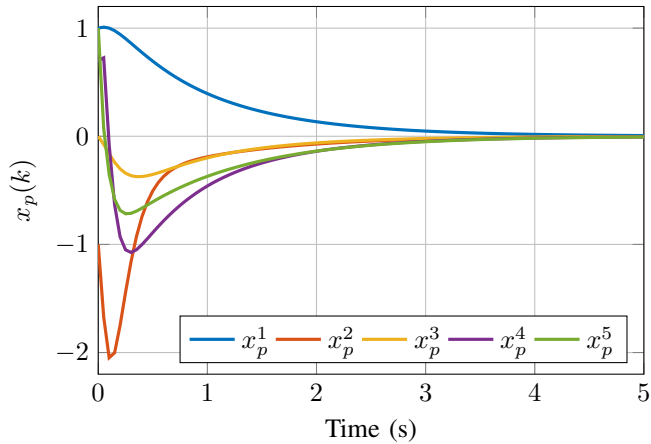}
	\caption{State of the plant controlled by the proposed encrypted controller when $1/L=2000$ and $1/s=10^4$.}
	\label{fig:xp}
\end{figure}

Fig.~\ref{fig:simul} depicts the performance error in terms of the difference between the control input of the original controller $u^\prime(k)$ and that of the encrypted controller $u(k)$,
under varying quantization parameters $1/L$ and $1/s$.
This demonstrates that the performance error can be maintained under a certain level, which tends to decrease
as either $1/L$ or $1/s$ increases.
Fig.~\ref{fig:xp} shows the state $x_p(k)=[x_p^1(k),\,x_p^2(k),\,\ldots,\,x_p^5(k)]^\top$ of the plant \eqref{eq:simulplant} equipped with the proposed encrypted controller
when $1/L=2000$ and $1/s=10^4$.
It can be seen that each element of the state approaches to zero as expected.

The average elapsed time for a control period---the time taken from the sensor to the actuator at each time step---was $0.0104$s, which is within the sampling period $0.05$s.
The experiment to measure the elapsed time was taken for $k\in [0,100)$.
Note that during one control period,
the encrypted controller performs $10$-homomorphic multiplications, since the order of the controller is $5$.

The time taken for each operation was approximately as follows: $510\mu$s for a single homomorphic multiplication, $1.5$ms for encrypting a plaintext, and $1$ms each for packing and unpacking procedures.
The time taken for decryption and homomorphic addition was usually insignificant.
Though the computation time for encryption, packing, and unpacking is longer than that of a single homomorphic multiplication, each of them is executed at most $2$ times during each control period, regardless of the dimension of the controller.
In contrast, the proposed encrypted controller executes $2n$-homomorphic multiplications at each time step, as shown in Table~\ref{table}.
All of the experiments were conducted using $2.9$GHz Intel Core i7-10700 CPU with $16$GB RAM.
This demonstrates the practicality of the proposed design.

\section{Conclusion} \label{sec:conclude}

We have presented an encrypted controller design
which does not involve infinitely many recursive homomorphic operations.
It is implementable through most HE schemes, regardless of somewhat, leveled fully, or fully HE.
The design is based on representing the controller output into a linear combination of a fixed number of previous inputs and outputs.
Furthermore, it is customized for RLWE-based cryptosystems, where a vector of messages can be encrypted into a single ciphertext and operated at once.
The efficiency of using this customized method, in terms of computation and communication, is discussed through numerical analysis.


\section*{Appendix}

\subsection{Packing and Unpacking}\label{appendix:packing}

	Suppose that $p$ is a power of $2$ and $N$ is a prime such that $N=1\,\,\,\mathrm{mod}\,\,2p$.
	Let $\zeta$ be the primitive $2p$-th root of unity modulo $N$, \textit{i.e.}, $\zeta^{2p}\,\,\,\mathrm{mod}\,\,N=1$ and $\zeta^k\,\,\,\mathrm{mod}\,\,N\neq 1$ for $k=1,\,2,\,\ldots,\,2p-1$.
	For $i=1,\,2,\,\ldots,\,p$, let $\zeta_i:=\zeta^{2i-1}\,\,\,\mathrm{mod}\,\,N$,
	whose multiplicative inverse is $\zeta_i^{-1}=\zeta^{2p-(2i-1)}\,\,\,\mathrm{mod}\,\,N$.
	We first construct a Vandermonde matrix as
	\begin{align*}
		\Theta:=\begin{bmatrix}
			1  & \zeta_1 & \cdots & \zeta_1^{p-1}\\
			1 & \zeta_2 & \cdots & \zeta_2^{p-1}\\
			\vdots & \vdots & \cdots & \vdots\\
			1 & \zeta_p & \cdots & \zeta_p^{p-1}
		\end{bmatrix}.
	\end{align*}
	Note that the inverse matrix of $\Theta$ is
	\begin{align*}
		\Theta^{-1}=p^{-1}\begin{bmatrix}
			1  & 1 & \cdots & 1\\
			\zeta_1^{-1} & \zeta_2^{-1} & \cdots & \zeta_p^{-1}\\
			\vdots & \vdots & \cdots & \vdots\\
			\zeta_1^{-(p-1)} & \zeta_2^{-(p-1)} & \cdots & \zeta_p^{-(p-1)}
		\end{bmatrix},
	\end{align*}
	where $p^{-1}$ is the multiplicative inverse of $p$ in $\Z_N$ such that $pp^{-1}\,\,\,\mathrm{mod}\,\,N=1$.
	Then, the packing and unpacking functions are defined as follows \cite{heaan}:
	\begin{align*}
		&\Pack\left(z\right):=\begin{bmatrix}
			1 & X & \cdots & X^{p-1}
		\end{bmatrix}\Theta^{-1}z\,\,\,\mathrm{mod}\,\,N,\\
		&\Unpack\left(a(X)\right):=\col\left\lbrace a\left( \zeta_i\right) \right\rbrace_{i=1}^p\,\,\,\mathrm{mod}\,\,N.
	\end{align*}
	
	\textit{Example:} Consider the case when $p=4$ and $N=17$.
	It is easily seen that the primitive eighth root of unity modulo $17$ is $2$,
	and the multiplicative inverse of $p$ in $\Z_{17}$ is $-4$.
    Then, the matrices $\Theta$ and $\Theta^{-1}$ are computed as
	\begin{align*}
		\Theta = \begin{bmatrix}
			1 & 2 & 4 & 8\\
			1 & 8 & -4 & 2 \\
			1 & -2 & 4 & -8\\
			1 & -8 & -4 & -2
		\end{bmatrix},\,
		\Theta^{-1}=
		\begin{bmatrix}
			-4 & -4 & -4 & -4\\
			-2 & 8 & 2 & -8\\
			-1 & 1 & -1 & 1\\
			8 & -2 & -8 & 2
		\end{bmatrix}.
	\end{align*}
	Given two vectors in $\Z_{17}^4$ as $u=\left[1,\,3,\,5,\,7\right]^\top$ and $v=\left[2,\,-4,\,-6,\,8\right]^\top$, one can obtain polynomials in $R_{4,17}$ as
	\begin{align*}
		\Pack(u)&=-7X^3+4X^2-7X+4\quad \text{and}\\
		\Pack(v)&=3X^3+8X^2+7X.
	\end{align*}
	The addition of the polynomials yields
	\begin{equation*}
		\Pack(u)+\Pack(v)\,\,\,\mathrm{mod}\,\,17=-4X^3-5X^2+4,
	\end{equation*}
	which is unpacked to $\left[3,\,-1,\,-1,\,-2\right]^\top=u+v\,\,\,\mathrm{mod}\,\,17$.
	Similarly, multiplying the polynomials over $R_{4,17}$ gives
	\begin{align*}
		&\Pack(u)\Pack(v)\,\,\,\mathrm{mod}\,\left(17,X^4+1\right)\\
		&= -4X^6+7X^5-4X^4+X^3-6X \,\,\,\mathrm{mod}\,\left(X^4+1\right)\\
		&= X^3+4X^2+4X+4,
	\end{align*}
	and the resulting polynomial is unpacked to $\left[2,\,5,\,4,\,5\right]^\top=u\circ v\,\,\,\mathrm{mod}\,\,17$.
	
\subsection{BGV scheme \cite{bgv}}\label{appendix:bgv}

The encryption, decryption, and basic homomorphic operations of the BGV scheme are defined as follows.
\begin{itemize}
	\item \textit{Parameters} ($N,\,p,\,q,\,\sigma$): Let $N\in\N$ and $q\in\N$ be coprime, where $q \gg N$, and $p$ be a power of $2$.
	Sampling $e\in R_{p,q}$ from the distribution $\chi$, which is denoted by $e\gets \chi$, indicates that each coefficient of $e$ is sampled from the discrete Gaussian distribution $N(0,\sigma)$.
	\item \textit{Secret key generation}: $\mathsf{sk}\gets \chi$.
	\item \textit{Encryption}: Sample $a\in R_{p,q}$ uniformly from $R_{p,q}$ and $e\gets \chi$.
	For a plaintext $m\in R_{p,N}$,
	\begin{align*}
		\Enc(m):=\begin{bmatrix}
			a\cdot \mathsf{sk}+Ne+m\\
			-a
		\end{bmatrix}\!\!\!\!\!\!\mod \!(X^p+1,q)\in R_{p,q}^2.
	\end{align*}
	\item \textit{Decryption}: For a ciphertext $\mathbf{c}\in R_{p,q}^\mathsf{n}$,
 $\Dec(\mathbf{c}):=\langle \mathbf{c},\, \mathbf{sk}\rangle \!\!\!\!\mod (X^p+1,q)\!\!\!\!\mod N \in R_{p,N},$
	where $\mathbf{sk}:=[1,\,\mathsf{sk},\,\ldots,\,\mathsf{sk}^\mathsf{n}]^\top\!\!\!\!\mod (X^p+1,q) $.
	\item \textit{Addition}: For $\mathbf{c}_1\in R_{p,q}^\mathsf{n}$ and $\mathbf{c}_2\in R_{p,q}^\mathsf{n}$, $\mathbf{c}_1\oplus \mathbf{c}_2:=\mathbf{c}_1+\mathbf{c}_2\!\!\!\!\mod q\in R_{p,q}^\mathsf{n}.$
	\item \textit{Multiplication}: For $\mathbf{c}_1\in R_{p,q}^2$ and $\mathbf{c}_2\in R_{p,q}^2$,
	\begin{align*}
		\Mult(\mathbf{c}_1,\mathbf{c}_2):=\begin{bmatrix}
			c_{1,1}c_{2,1} \\
			c_{1,1}c_{2,2}+c_{1,2}c_{2,1}\\
			c_{1,2}c_{2,2}
		\end{bmatrix}\!\!\!\!\!\mod (X^p+1,q),
	\end{align*}
	where $\mathbf{c}_1=:[c_{1,1},\,c_{1,2}]^\top$ and $\mathbf{c}_2=:[c_{2,1},\,c_{2,2}]^\top$.
\end{itemize}

\bibliography{refs}
\bibliographystyle{IEEEtran}



\vspace{-11pt}
\begin{IEEEbiography}[{\includegraphics[width=1in,height=1.25in,clip,keepaspectratio]{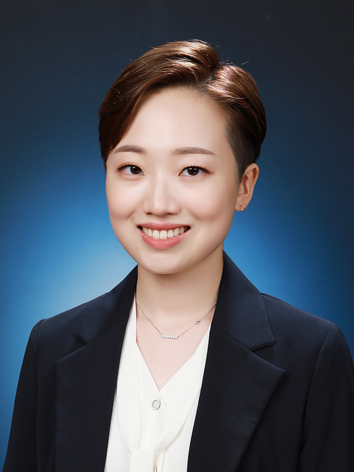}}]{Joowon Lee} received the B.S. degree in electrical and computer engineering in 2019, from Seoul National University, South Korea.
She is currently a combined M.S./Ph.D. student in electrical and computer engineering at Seoul National University, South Korea. 
Her research interests include encrypted control systems and data-driven control.
\end{IEEEbiography}

\begin{IEEEbiography}[{\includegraphics[width=1in,height=1.25in,clip,keepaspectratio]{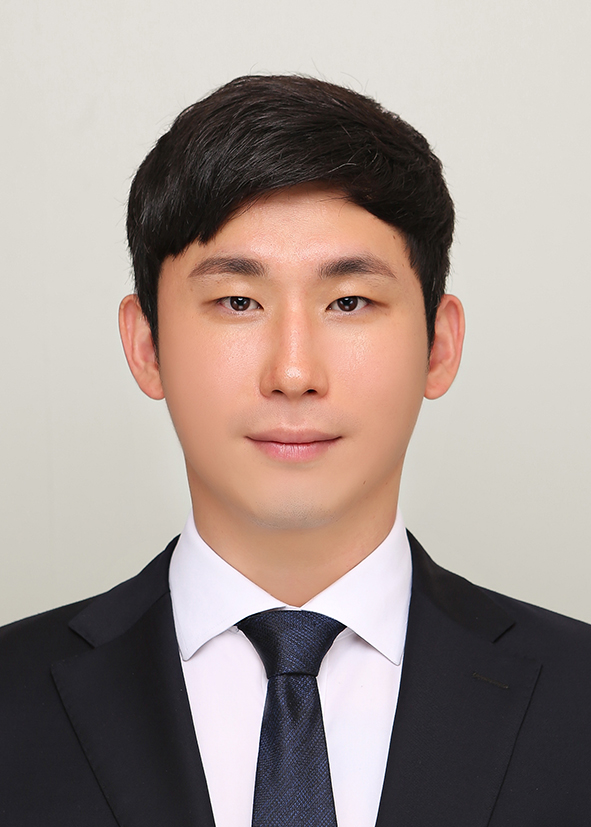}}]{Donggil Lee} received the B.S. and Ph.D. degrees from the Department of Electrical Engineering and Computer Science from Seoul National University, Korea, in 2015 and 2023, respectively. He served as a postdoctoral researcher at the Korea Institute of Science and Technology until 2024. Since then, he has been an Assistant Professor in the Department of Electrical Engineering at Incheon National University, Korea. His research interests focus on various aspects of multi-agent systems, including distributed estimation and control, encrypted control, and task allocation.
\end{IEEEbiography}

\begin{IEEEbiography}[{\includegraphics[width=1in,height=1.25in,clip,keepaspectratio]{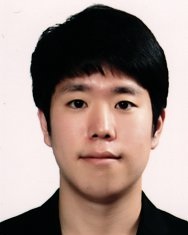}}]{Junsoo Kim} received the B.S. degrees in electrical engineering and mathematical sciences in
 2014, and the M.S. and Ph.D. degrees in electrical engineering in 2020, from Seoul National University, South Korea, respectively. 
 He held the Postdoc position at KTH Royal Institute of Technology, Sweden, till 2022. 
  He is currently an
 Assistant Professor at the Department of Electrical and Information Engineering, Seoul National University of Science and Technology, South Korea. His research interests include security
 problems in networked control systems and encrypted control systems.
\end{IEEEbiography}

\begin{IEEEbiography}[{\includegraphics[width=1in,height=1.25in,clip,keepaspectratio]{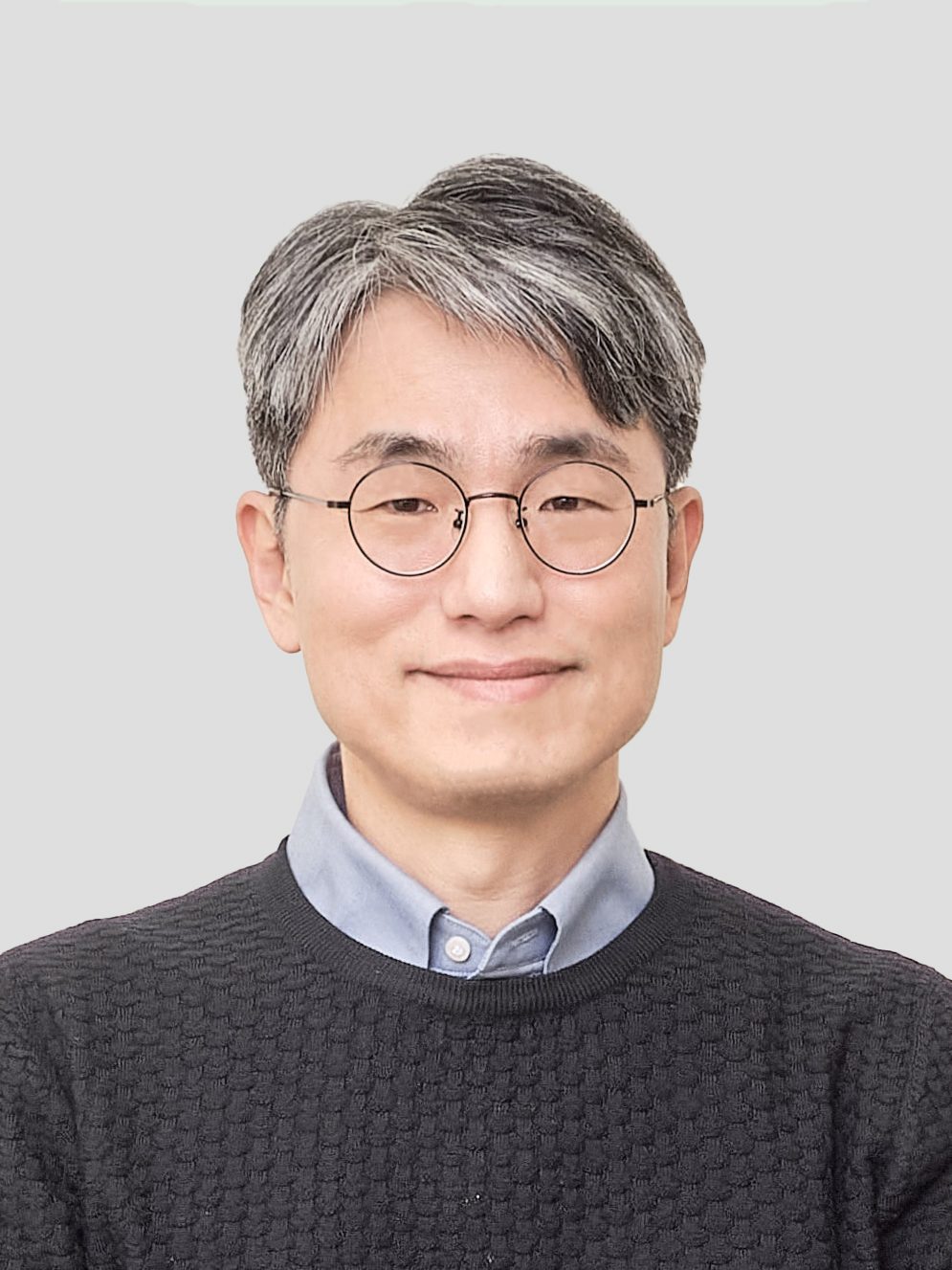}}]{Hyungbo Shim}
received his B.S., M.S., and Ph.D. degrees from Seoul National University, Korea, and held the post-doc position at University of California, Santa Barbara till 2001.
He joined Hanyang University, Seoul, in 2002.
Since 2003, he has been with Seoul National University, Korea.
He served as an associate editor for Automatica, IEEE Transactions on Automatic Control, International Journal of Robust and Nonlinear Control, and European Journal of Control, and as an editor for International Journal of Control, Automation, and Systems.
He serves for the IFAC World Congress 2026 as the general chair.
His research interests include stability analysis of nonlinear systems, observer design, disturbance observer technique, secure control systems, and synchronization for multi-agent systems.
\end{IEEEbiography}

\vfill

\end{document}